\documentclass[12pt,preprint]{aastex}
\usepackage{natbib}
\usepackage{amsmath}

\begin{document}
\bibliographystyle{plainnat}
\title{Selection effects in Gamma Ray Bursts correlations: consequences on the ratio between GRB and star formation rates}

\author{Dainotti, M. G. \altaffilmark{1,2,3}, Del Vecchio, R. \altaffilmark{3}, Nagataki, S. \altaffilmark{1}, Capozziello, S. \altaffilmark{4,5,6}}

\altaffiltext{1}{Astrophysical Big Bang Laboratory, Riken, 2-1 Hirosawa, Wako, Saitama 351-0198, Japan, maria.dainotti@riken.jp.}
\altaffiltext{2}{Physics Department, Stanford University, Via Pueblo Mall 382, Stanford, CA, USA, E-mail: mdainott@stanford.edu}
\altaffiltext{3}{Obserwatorium Astronomiczne, Uniwersytet Jagiello\'nski, ul. Orla 171, 31-501 Krak{\'o}w, E-mails: delvecchioroberta@hotmail.it, dainotti@oa.uj.edu.pl, mariagiovannadainotti@yahoo.it}
\altaffiltext{4}{Dipartimento di Fisica, Universit$\acute{a}$ di Napoli "Federico II", Compl. Univ. di Monte S. Angelo, Ediﬁcio G, Via Cinthia, I-80126 Napoli, Italy, E-mail: capozziello@na.infn.it}
\altaffiltext{5}{INFN Sez. di Napoli, Complesso Universitario di Monte S. Angelo, Via Cinthia, Ediﬁcio N, 80126 Napoli, Italy}
\altaffiltext{6}{Gran Sasso Science Institute (INFN), viale F. Crispi 7, I-67100 L'Aquila, Italy}

\begin{abstract}
 Gamma Ray Bursts (GRBs) visible up to very high redshift have become attractive targets as 
potential new distance indicators. It is still not clear whether the relations proposed so far originate from an unknown GRB physics or result from selection effects. We
investigate this issue in the case of the $L_X-T^*_a$ correlation (hereafter LT) between the X-ray luminosity $L_X (T_a)$ at 
the end of the plateau phase, $T_a$, and the rest frame time $T^{*}_a$. 
We devise a general method to build mock data sets starting from a GRB world model and taking
into account selection effects on both time and luminosity. This method 
shows how not knowing the efficiency function could influence the evaluation of the intrinsic slope of any 
correlation and the GRB density rate. We investigate
 biases (small offsets in slope or normalization) that would occur
in the LT relation as a result of truncations, possibly present in the intrinsic 
distributions of $L_X$ and $T^*_a$. We compare these results with the ones 
in Dainotti et al. (2013) showing that in both
cases the intrinsic slope of the LT correlation is $\approx -1.0$. This method is general, therefore relevant to investigate if any other GRB correlation is generated by the biases themselves. 
Moreover, because the farthest GRBs and star-forming galaxies probe the reionization epoch, we
evaluate the redshift-dependent ratio $\Psi(z)=(1+z)^{\alpha}$ of the GRB rate to star formation rate (SFR). We 
found a modest evolution $-0.2\leq \alpha \leq 0.5$ consistent with Swift GRB afterglow plateau in the 
redshift range $0.99<z<9.4$.
\end{abstract}
\keywords{stars: gamma-ray burst: general, statistics, methods: data analysis.}

\section{INTRODUCTION}
GRBs are the farthest sources, seen up to redshift $z = 9.46$ \citep{Cucchiara2011}, and
if emitting isotropically they are also the most powerful, (with $E_{iso} \le 10^{54} erg \ s^{-1}$), 
objects in the Universe. Notwithstanding the variety of their peculiarities, some common features may be identified by looking at their 
light curves. A crucial breakthrough in this area has been the observation of GRBs by the Swift satellite, launched 
in 2004. With the instruments on board, the Burst Alert Telescope (BAT, 15-150 keV), the X-Ray Telescope 
(XRT, 0.3-10 keV), and the Ultra-Violet/Optical Telescope (UVOT, 170-650 nm), Swift provides a rapid follow-up of the
afterglows in several wavelengths with better coverage than previous missions. Swift observations have revealed a 
more complex behavior of the light curves afterglow \citep{Obrien2006,Sakamoto2007} that can be divided 
into two, three and even more segments in the afterglows. The second segment, when it is flat, is called plateau 
emission. A significant step forward in determining common features in the afterglow light curves was made by fitting 
them with an analytical expression \citep{Willingale2007}, called hereafter W07.
 This provides the opportunity to look for universal features that could provide a redshift
independent measure of the distance from the GRB, as in studies of correlations between GRB isotropic
energy and peak photon energy of the $\nu F_{\nu}$ spectrum, $E_{iso}-E_{peak}$, \citep{Lloyd1999,Amati2009}, the beamed total 
energy $E_{\gamma}-E_{peak}$ \citep{Ghirlanda2004,Ghirlanda2006},
the Luminosity-Variability, L-V \citep{Norris2000,Fenimore2000}, L-$E_{peak}$ \citep{Yonetoku2004} and Luminosity- $\tau$ lag \citep{Schaefer2003}.\\
Dainotti et al. (2008, 2010), using the W07 phenomenological law for the light curves 
of long GRBs, discovered a formal anti-correlation between the X-ray luminosity at the end of the plateau 
$L_X$ and the rest frame plateau end-time, $T^*_a = T^{obs}_a/(1 + z)$, where $T^*_a$ is in seconds and $L_X$ is in erg/s. The normalization and the slope parameters 
$a$ and $b$ are constants obtained by the D'Agostini fitting method \citep{Dagostini2005}. Dainotti
et al. (2011a) attempted to use the LT correlation as a possible redshift estimator, but the
paucity of the data and the scatter prevents from a definite conclusion at least for a sample
of 62 GRBs. In addition, a further step to better understand the role of the plateau emission
has been made with the discovery of new significant correlations between $L_X$, and the mean
luminosities of the prompt emission, $<L_{\gamma,prompt}>$ \citep{Dainotti2011b}.
 The LT anticorrelation is also a useful test for theoretical models such as the accretion
models, \citep{Cannizzo2009,Cannizzo2011}, the magnetar models \citep{Dallosso2011,Bernardini2011, 
Bernardini2012,Rowlinson2010,Rowlinson2013,Rowlinson2014}, the prior emission model \citep{Yamazaki2009}, 
the unified GRB and AGN model \citep{Nemmen2012} and the fireshell model \citep{Izzo2012}. Moreover, \cite{hascoet2014} and \cite{vanerten14b} consider both the LT and the $L_X$-$<L_{\gamma,prompt}>$ correlation to discriminate among several models proposed for the origin of plateau. In \cite{leventis2014} and in \cite{vanerten14a} a smooth energy injection through the reverse shock has been presented as a plausible explanation for the origin of the LT correlation. 
Furthermore, also other authors were able to reproduce and use the LT correlation to extend it in the optical band \citep{Ghisellini2009}, to extrapolated it into correlations of the prompt emission \citep{Sultana2012} and to use the same methodology to build an analogous correlation in the prompt \citep{Qi2012}.
Finally, it has been applied as a cosmological tool \citep{Cardone2009,Cardone2010,Dainotti2013b,Postnikov2014}.
Impacts of detector thresholds on cosmological standard candles have also been considered \citep{Shahmoradi2009,Petrosian1998,Petrosian1999,Petrosian2002,Cabrera2007}.
However, because of large dispersion \citep{Butler2010,Yu2009} and absence of good calibration none of these
correlations allow the use of GRBs as good standard candles as it has been done e.g. with type Ia Supernovae.
 An important statistical technique to study selection effects for treating data truncation in GRB correlations 
is the \cite{Efron1992} method. Another way to study the same problem in GRB correlations, derived modeling the high-energy properties of GRBs, have been reported in Butler et al. (2010). In the latter paper it has been shown that well-known examples of these correlations have common features indicative of strong contamination by selection effects. We compare this procedure with the method introduced by Efron \& Petrosian (1992) and applied to the LT correlation \citep{Dainotti2013a}.
The paper is organized as follows: section \ref{GRB and SRF ratio} introduces the relation between GRB and SFR, section \ref{GRB world model} is dedicated to the analysis of a GRB scaling relation, in particular we consider the LT correlation as an example, but the procedure described can be adopted for any other correlation. 
In section \ref{simulating the sample} we describe how to build the GRB samples, in section \ref{redshift evolution} we analyze
the redshift evolution of the slope and normalization of the LT correlation. In section \ref{impact of selection effects} we study the selection effects related to 
simulated samples assuming different normalization and slope values. Then, in section \ref{Conclusions} we draw 
conclusions on the intrinsic slope of the LT correlation and on the evaluation of the redshift-dependent ratio between GRB and star formation rates.

\section{The relation between GRB rate and the star formation rate}\label{GRB and SRF ratio}

In order to understand the relation between GRBs and the
star formation, it is often assumed that the GRB
rate (RGRB) is proportional to the SFR then the
predicted distribution of the GRB redshift is compared to the observed distribution \citep{Totani97,Mao98,wijers98,porciani01,natarajan05,
jakobsson06,daigne07,le07,coward07,mao10}. However, this relationship is not an easy task to handle, because some studies show that GRBs do not seem to trace the star formation unbiasedly \citep{Lloyd1999}. Namely, the ratio between the RGRB and SFR, RGRB/SFR, increases with redshift \citep{Kistler2008,Yuksel2007}
significantly. This means that GRBs are more frequent for a given star formation rate density
at earlier times. In fact, while observations
consistently show that the comoving rate density of
star formation is nearly constant in the interval $1 \leq z \leq 4$
\citep{Hopkins2006}, the comoving rate density of
GRBs appears evolving distinctly. In our approach we explicitly take into consideration this issue when we fit the observed GRB rate with the model.
Selection effects involved in a GRB sample are of two kinds : the GRB
detection and localization; and the redshift determination
through spectroscopy and photometry of the GRB afterglow or
the host galaxy. These problems have been object of extensive study in literature \citep{Bloom2003,Fiore2007,Guetta2007}. Moreover, the Swift trigger, is very complex and the sensitivity of the detector is very difficult to parameterize exactly \citep{Band2006}, but in this case not dealing with prompt peak energy we do not have to take into consideration the double truncation present in data \citep{Lloyd1999}. In the case of plateau it is easier, since an effective luminosity
threshold appears to be present in the data which can be approximated by a $0.3-10$ keV energy flux limit $F_{lim}\equiv 2 \times 10^{−12}$ erg $cm^{−2}$ $s^{−1}$ \citep{Dainotti2013a}. The luminosity threshold is then $L_{lim} = 4\pi D_L^2(z,\Omega_M,H_0) F_{lim}$, where $D_L$ is the luminosity distance to the burst. Throughout the paper, we assume a flat universe with $\Omega_M=0.28$, $\Omega_{\lambda}=0.72$ and $H_0$=70 km $s^{−1}$ $Mpc^{−1}$.
In our approach below several models are considered and then the one that best matches the GRB rate with star formation rate has been chosen.

\subsection{GRBS WORLD MODEL}\label{GRB world model}

 We derive a model capable of reproducing the observed Swift GRB rate as a function of redshift, 
luminosity and time of the plateau emission.\\
Rest frame time and luminosity at the end of the GRB plateau emission show strong correlations as discovered by \cite{Dainotti2008}
and later updated by \cite{Dainotti2010a,Dainotti2011a,Dainotti2011b,Dainotti2013b}. Therefore, all these quantities must be considered in deriving reliable rates. 
We characterize the GRB rate as a product of terms involving the redshift z of the bursts, the isotropic equivalent 
luminosity release (0.3-10 keV) $L_X$ and the duration $T^{*}_a$.

\begin{figure}
\centering
\includegraphics[width=0.50\hsize,angle=0,clip]{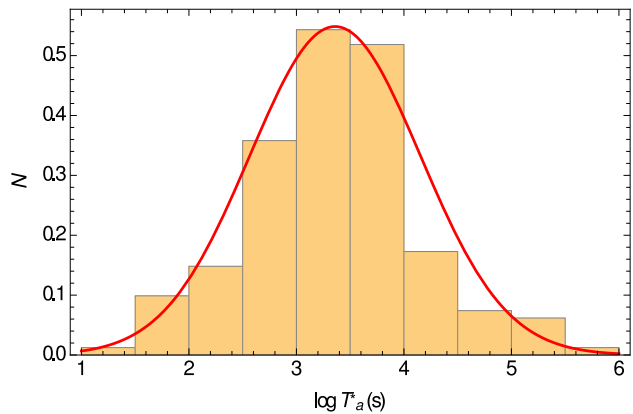}
   \caption{\footnotesize Probability Density Distribution of $T^*_a$, the rest frame end time of the plateau, for GRBs 
   observed from 2005 January until 2014 July analyzed following the Dainotti et al. (2013a) approach with a 
   superimposed best fit of the Gaussian distribution.}
   \label{fig:6}
\end{figure}

\begin{figure}
\centering
\includegraphics[width=0.50\hsize,angle=0,clip]{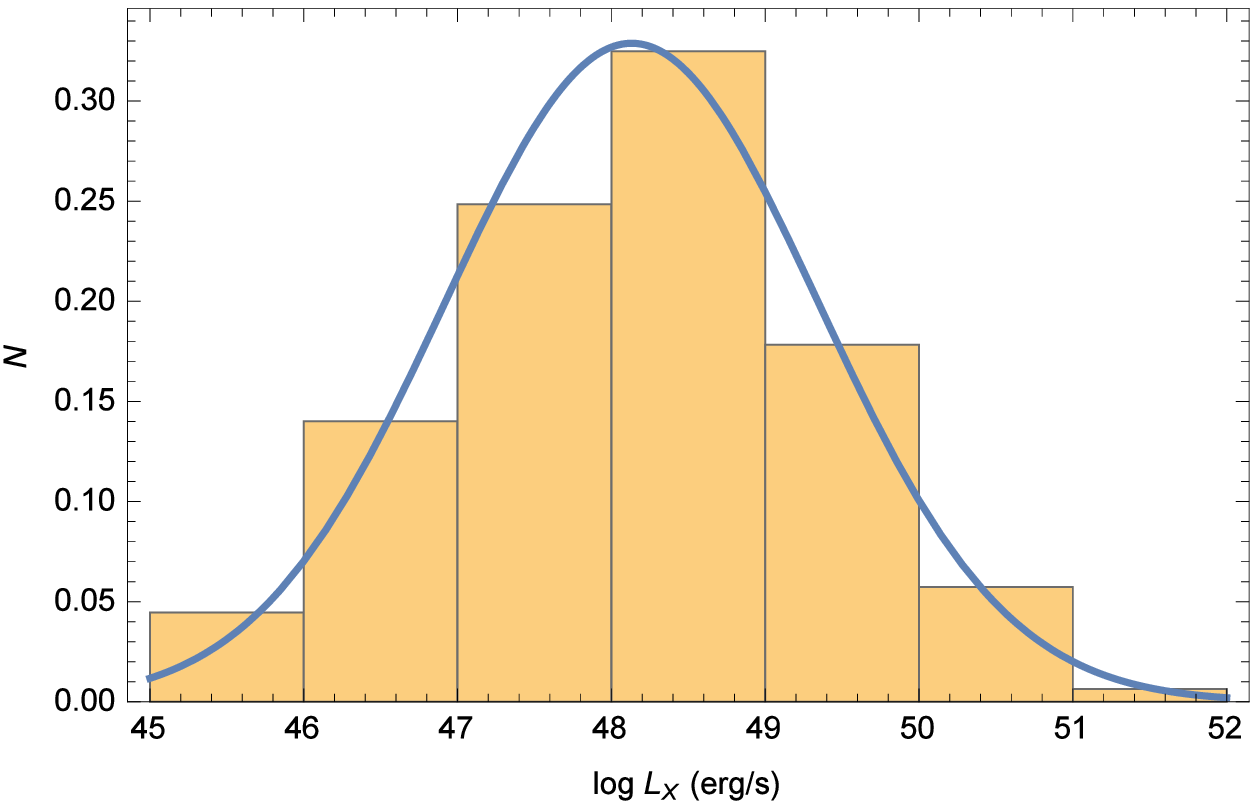}
   \caption{\footnotesize Probability Density Distribution of $L_X(T_a)$ at the end of the plateau for GRBs
   observed from  2005 January until 2014 July analyzed following the Dainotti et al. (2013a) approach with a 
   superimposed best fit of the Gaussian distribution.}
   \label{fig:7}
\end{figure}

 Let us assume that a scaling relation exists so that the luminosity $L_X(T_a)$ for a GRB with time scale 
$T^*_a$ at redshift z is given by :

\begin{equation}
  \lambda=\alpha_0+\alpha_{\tau}\tau + \alpha_{\zeta}\zeta
\label{scaling}
\end{equation}

 where we have introduced the compact notation

\begin{equation}
\begin{cases} 
\lambda= log \ L_X(T^*_a) \\ \tau=log\ [T^*_a/(1+z)] \\ \zeta=log\ (1+z).
\end{cases} 
\label{lambda}
\end{equation}

 and the term $\zeta$ accounts for redshift evolution. The luminosity is normalized by the unit of 1 erg $s^{-1}$ and the time by
the unit of 1 s, so that non dimensional quantities are considered. All the observables in this model are computed in the rest frame, because we are testing the role played by selection effects in the rest frame, being the LT correlation rest frame corrected.
 Independently, on the physical interpretation of this relation, (in fact, there are several models that can 
reproduce it as we have mentioned in the introduction) we can nevertheless expect GRBs to follow Equation 
\ref{scaling} with a scatter $\sigma_{\lambda}$. 
Moreover, the zero point $\alpha_0$ may be known only up to a given uncertainty $\sigma_{\alpha}$. Following the 
approach of Butler et al. (2010) applied for prompt correlations, we assume that $\lambda$ can be approximated by 
Gaussian distribution with mean $\lambda_0$, expressed in Equation \ref{scaling}, and the variance $\sigma_{int}$ as 
the intrinsic scatter of the correlation. 
We also write the probability that a GRB with given ($\tau$, $\zeta$) values has a luminosity $\lambda$ as follows:

\begin{equation}
P_{\lambda}(\lambda,\tau,\zeta) \propto exp[-\frac{1}{2}[\frac{\lambda-(\alpha_0+\alpha_{\tau}\tau + \alpha_{\zeta}\zeta)}{\sigma_{int}}]^2]
\label{plambda}
\end{equation}

 with $\sigma_{int}^2 = \sigma_{\lambda}^2 + \sigma_{\alpha}^2$, with $\sigma_{\alpha}$ the uncertainty 
of the $\alpha_0$ value and $\sigma_{\lambda}$ the uncertainty on the luminosity value.\\
The approximation of a Gaussian distribution both for the luminosity and time 
is motivated by the goodness of the fit which gives a probability $P=0.46$ and $P=0.61$ respectively, see Fig. \ref{fig:6} and \ref{fig:7}. We note that the mean, (indicated with $< >$) $<T_a>=3.35$ (s)
with a variance $\sigma_{T_a}=0.77$ (s) and $<L_X>=48.04$ (erg/s) with a variance $\sigma_{L_X}=1.37$ (erg/s) are represented respectively in Fig. \ref{fig:6} and \ref{fig:7}.

 In order to obtain the number of GRBs with a given luminosity $\lambda$, we need to integrate over the 
distributions of $\tau$ and $\zeta$. We will assume, for simplicity, that $\tau$ follows a truncated Gaussian 
law.

\begin{equation}
 P_{\tau}(\tau) \propto \begin{cases} exp[-\frac{1}{2}(\frac{\tau-\tau_0}{\sigma_{\tau}})^2] & \tau_L<\tau<\tau_U \\ 0 & \tau \le \tau_L \ or \ \tau \ge \tau_U
\end{cases}
\label{ptau}
\end{equation}

 where $\tau_L$ and $\tau_U$ indicate respectively the lower limit and upper limit of the observed $\tau$ 
distribution and $\tau_0$ is the mean value of this distribution. The limits of $\tau$ are taken from an updated 
sample of $T^*_a$ composed of 176 GRBs afterglows, with firm redshift determination, from January 2005 till July 2014. The analysis follows the criteria
adopted in Dainotti et al. (2013a).

If we assume that the GRBs trace the cosmic star formation rate, we can model their redshift distribution 
following Butler et al. (2010) as:

\begin{equation}
P_z(z) \propto \frac{\dot{\rho}_*(z)}{1+z}\frac{dV}{dz}
\label{pzeta}
\end{equation}

 where $\dot{\rho}_{*}(z)$ is the comoving
GRB rate density, V is the universal volume, and the factor $(1+z)$ accounts for cosmic time dilatation and

\begin{equation}
\frac{dV}{dz} \propto \frac{r^2(z)}{E(z)} 
\label{volume}
\end{equation}

 with $r(z)$ the comoving distance and $E(z)=H(z)/H_0$ the Hubble parameter normalized to its present day 
value.\\
 Collecting the different terms, we can finally write the true, detector-independent event $\mathcal{N}$ 
differential rate, for z, $\log T^*_a$ and $\log L_X$, as:

\begin{equation}
\frac{d\mathcal{N}}{d\lambda d\tau dz} \propto \Psi(z) P_{\lambda}(\lambda, \tau, \zeta) P_{\tau}(\tau)P_z(z).
\label{N1}
\end{equation}

We here note that we have introduced the term of the evolution in redshift, $\Psi(z)=(1+z)^{\alpha}$, following the approach of Lloyd \& Petrosian (1999), Dermer (2007) and Robertson \& Ellis (2012). In Dermer (2007) assuming that the emission properties of GRBs do not change with time, they find that the Swift data can only be fitted if the comoving rate density of GRB sources exhibits positive evolution to $z>~3-5$. In our approach we introduce evolution starting from $z \ge 0.99$.

 So using the above expression for $P_{\tau}$, we find that the number of GRBs with luminosity in 
the range ($\lambda$, $\lambda+d\lambda$) and redshift between $z$ and $z+dz$ is:

\begin{equation}
\frac{d\mathcal{N}}{d\lambda dz}\propto \Psi(z) \frac{\dot{\rho}_*(z)(dV/dz)}{1+z} \frac{\mathcal{F}_{\tau_U}-\mathcal{F}_{\tau_L}}{\sqrt{8\pi \sigma^2_{\tau}}}
  exp[-\frac{1}{2}[\frac{ \lambda-(\alpha_0+\alpha_{\tau}\tau + \alpha_{\zeta}\zeta)}
{\sqrt{\sigma^2_{int}}}]^2]
\label{N2}
\end{equation}

 where $\mathcal{F}_{\tau_U}$ and $\mathcal{F}_{\tau_L}$ are the error functions\footnote{We remind that the
usual definition of the error function is
\begin{equation}
erf(x)=\frac{2}{\sqrt{\pi}}\int^x_0 e^{-t^2}dt.  
\end{equation}} of the lower and upper limit of the time distribution.
 Note that Equation \ref{N2} is defined up to an overall normalization constant which can be solved by 
imposing that the integral of $d\mathcal{N}/d\lambda dz$ over ($\lambda$, $z$) gives the total number of observed 
GRBs. Actually, this is not known since we do not observe all GRBs, but only those passing a given set of selection 
criteria. However, we will be only interested in the fraction of GRBs in a cell in the 2D ($\lambda$, $z$) space so 
that we do not need this quantity.\\
We are aware that we don't map out the true LT relation given selection effects and the 
observed LT relation. Doing this would require modeling the selection of the GRB sample itself (using the 
gamma ray threshold) and also seeking to understand the tie between the GRB flux and the afterglow $L_X$. However, the relation between flux and $L_X$ has been already studied by Dainotti et al. (2013a) and reported briefly in the previous section. Here we computed the new limit related to the updated sample, as it has been shown in the middle panel of Fig \ref{fig:4}.

\section{SIMULATING THE GRB SAMPLES}\label{simulating the sample}
 The GRBs rate given by Equation \ref{N2} has been derived by implicitly assuming that all the GRBs can be 
detected notwithstanding their observable properties. This is actually not the case. As an example, we will consider 
hereafter the LT correlation although the formalism and the method we will develop can be easily extended to
whatever scaling law.
 For the LT case, there are two possible selection effects. First, each detector has an efficiency 
which is not the same for all the luminosities. Only GRBs with $\lambda > \lambda_L$, where $\lambda_L$ is the lowest 
detectable luminosity for a given instrument, can be detected while all the GRBs with $\lambda$ larger than a 
threshold luminosity $\lambda_U$ will be found.

\begin{figure}
\includegraphics[width=0.32\hsize,angle=0,clip]{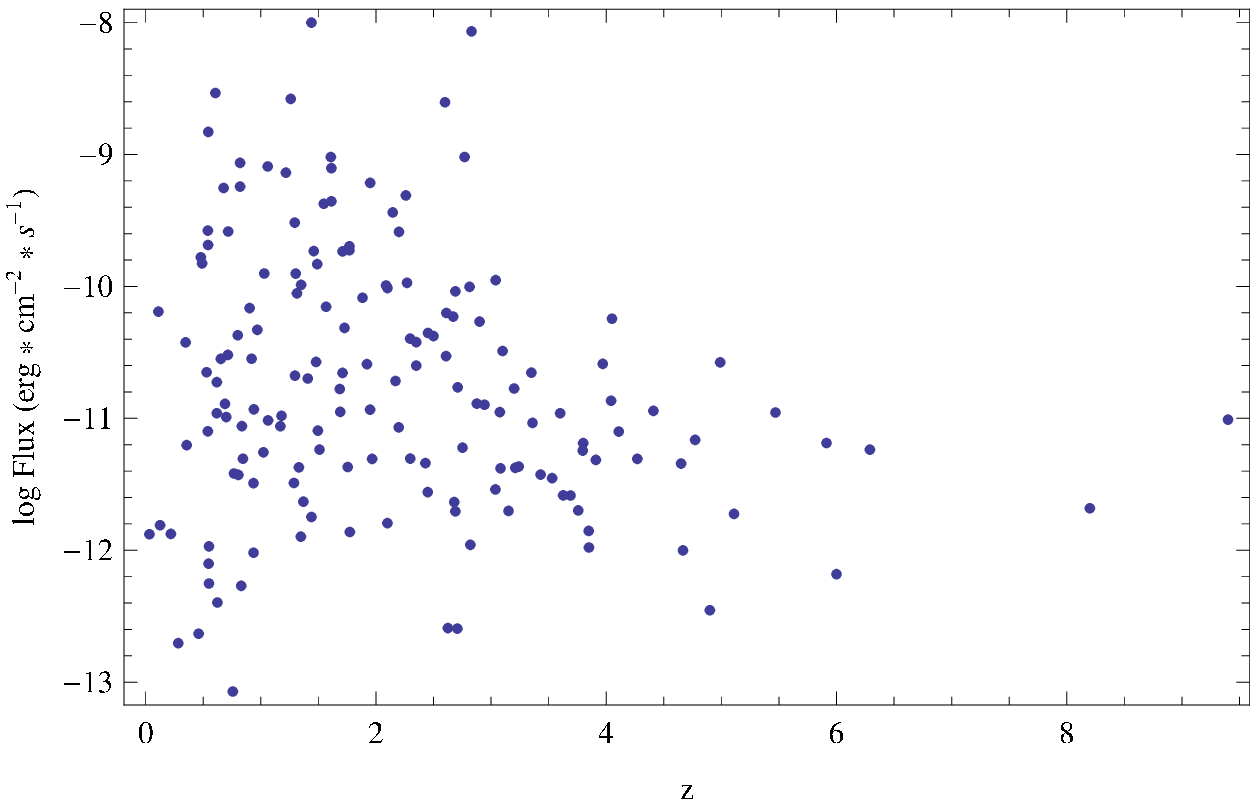}
\includegraphics[width=0.32\hsize,angle=0,clip]{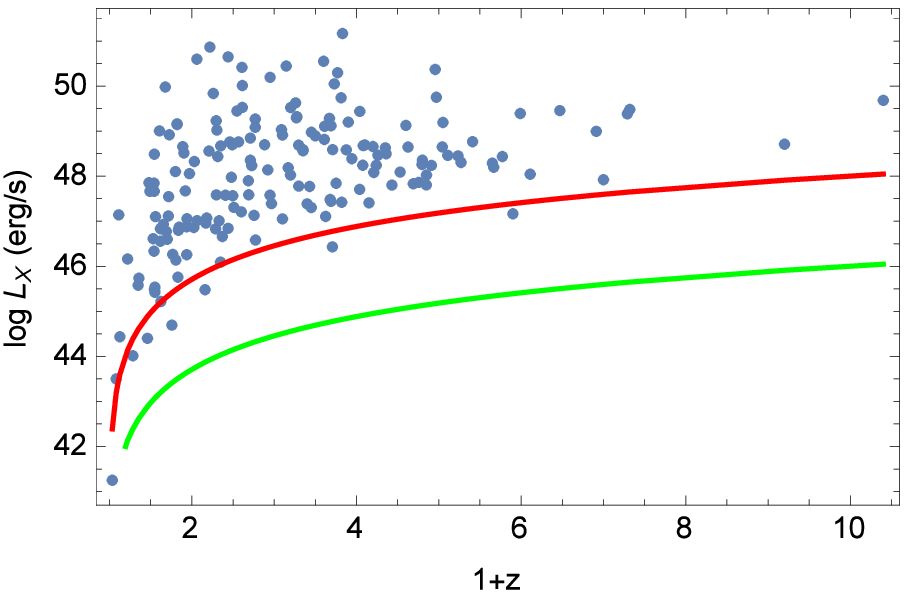}
\includegraphics[width=0.32\hsize,angle=0,clip]{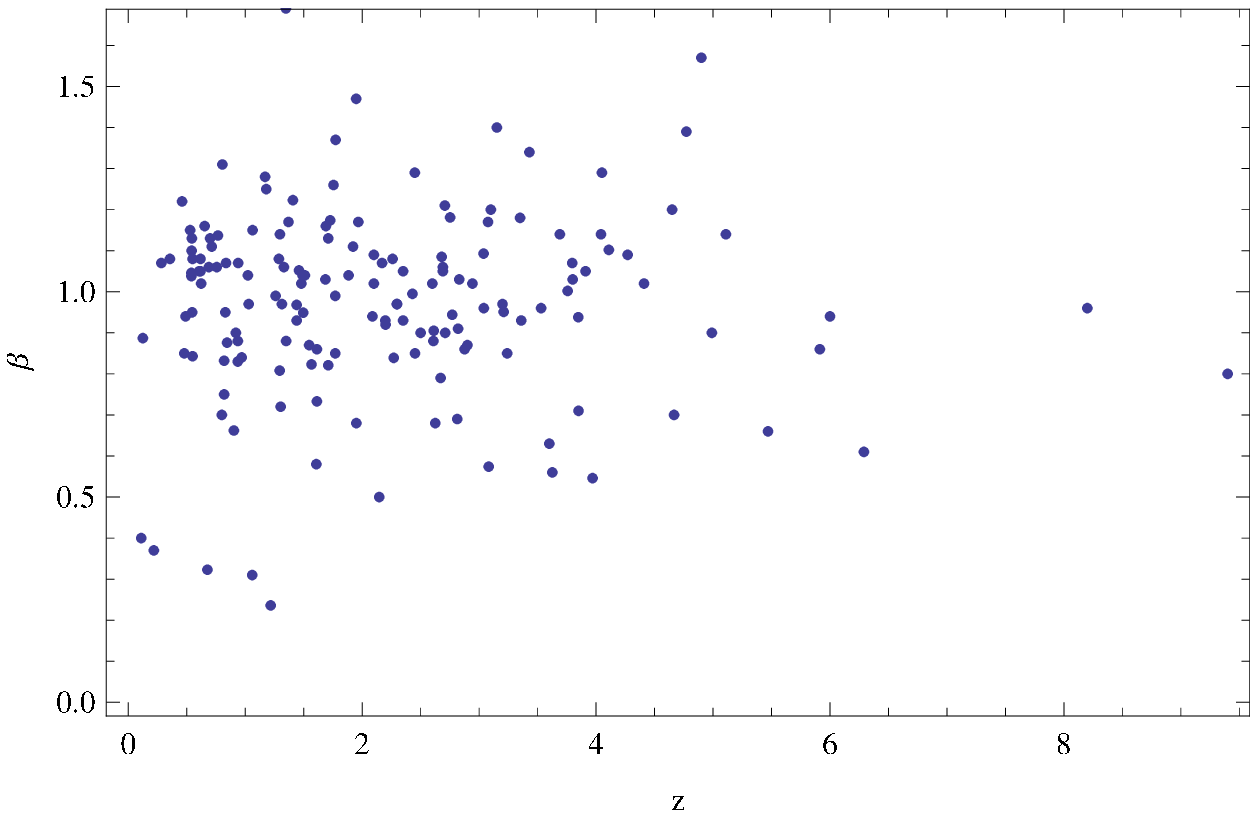}
\caption{\footnotesize Flux at the end of the plateau phase, Flux($T_a$), (left panel) and the spectral index, $\beta$, (right panel)
as a function of redshift. The limiting luminosity, $\log L_X$ vs $1+z$ shows (middle panel) two lines, one for the limiting flux, $F_{Swift,lim}=10^{-14}$ erg $cm^{-2}$ $s^{-1}$ and the other one is the most suitable for a plateau duration of $10^{4}$ s, which is $2 \times 10^{-12}$ erg $cm^{-2}$ $s^{-1}$.}
  \label{fig:4}
\end{figure}

 Moreover, it is likely that the efficiency of the detector is not constant, but is rather a function of the 
luminosity. We will therefore introduce an efficiency function $\mathcal{E}_{\lambda}(\lambda)$ whose functional 
expression is not known in advance, but can only take values in the range (0,1). A second selection effect is related 
to the time duration of the GRB. Indeed, in order to be included in the sample used to calibrate the LT 
correlation, the GRB afterglow has to be measured over a sufficiently long time scale to make possible to fit the data
and extract the relevant quantities. If $\tau$ is too small, as it has been shown in Dainotti et al. (2013a) the 
minimum rest frame time is 14 s, few points will be available for the fit, while, on the contrary, large $\tau$ 
values will give rise to afterglow light curves which could be well sampled by the data. Again, we can parametrize 
these effects introducing a second efficiency function $\mathcal{E}_{\tau}(\tau)$ so that the final observable rate 
is the following:

\begin{equation}
\frac{d\mathcal{N}_{obs}}{d\lambda dz} \propto \frac{d\mathcal{N}}{d\lambda dz}\times\mathcal{E}_{\lambda}(\lambda)\mathcal{E}_{\tau}(\tau).
\label{N3}
\end{equation}

We point out that our formulation, which takes into account of the efficiency functions $\mathcal{E}_{\lambda}(\lambda)$ and $\mathcal{E}_{\lambda}(\tau)$ in the final observed GRB rate is similar to the approach by Robertson \& Ellis (2012) in Equation 1, in which the additional factor K is presented. K  is 
equivalent to our $\mathcal{E}_{\lambda}(\lambda)$ and $\mathcal{E}_{\lambda}(\tau)$.

 It is worth noting that Equation \ref{N3} is actually still a simplified description. Indeed, 
it is in principle possible that other selection effects take place involving observable quantities not considered 
here, as for example $\beta$ and the redshift. However, these parameters enter in the determination of $\lambda$ so 
that one can (at least in a first order approximation) convert selection cuts on them in a single efficiency function 
depending only on $\lambda$ (for the dependence of the flux on the redshift see left panel of Fig. \ref{fig:4}).
However, as we can see from Fig. \ref{fig:4} $\beta$ is constant with redshift, and there is no correlation between 
those two quantities, in fact the Spearman correlation 
coefficient is $\rho=-0.062$. Nevertheless, Equation \ref{N3} provides a 
reasonably accurate description of the observable GRB rate.\\

 In order to evaluate Equation \ref{N3} there are different quantities to determine. First, we need to set 
the scaling coefficients ($\alpha_0$, $\alpha_{\tau}$, $\alpha_{\zeta}$) and the intrinsic scatter $\sigma_{int}$. 
Second, the mean and variance of the $\tau$ distribution ($\tau_0$,$\sigma_{\tau}$) has to be given. Finally, an 
expression for the cosmic SFR $\dot{\rho}_*(z)$ has to be assigned. None of these quantities is actually available. 
In principle, one could assume a SFR law and fit for the model parameters to a large enough GRBs sample with measured 
($\lambda$, $\tau$, $\zeta$) values. To this end, one should know the selection function 
$\mathcal{E}_{\lambda}(\lambda)\mathcal{E}_{\tau}(\tau)$ which is not the case.
Studies of how light curves would appear to a gamma-ray detector here on Earth have been performed \citep{kocevski2013}. In this paper the prompt emission pulses are investigated and the conclusion is that even a perfect detector that observes over a limited energy range would not faithfully measure the expected time dilation effects on a GRB pulse as a function of redshift.
 
\begin{figure}[htbp]
\centering
\includegraphics[width=0.49\hsize,angle=0,clip]{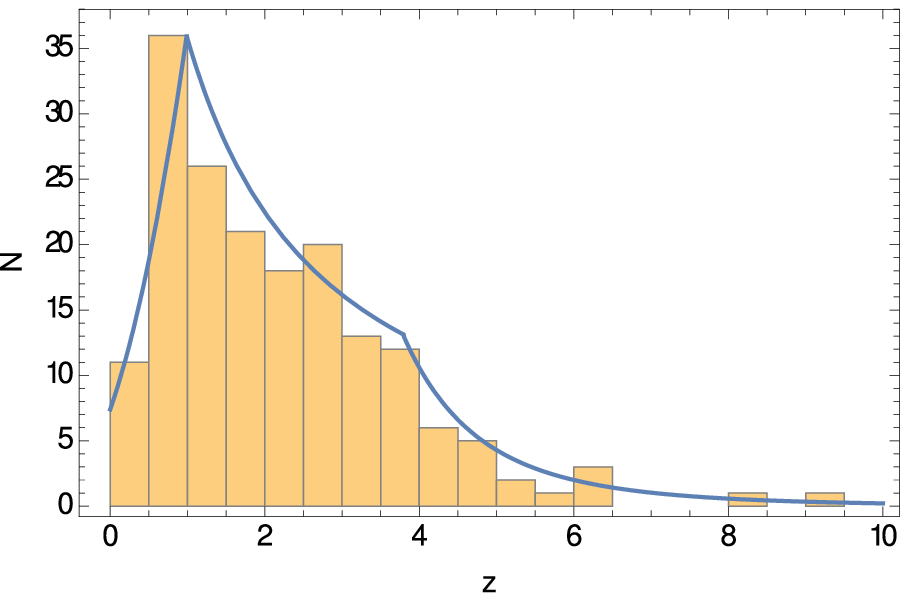}
\includegraphics[width=0.49\hsize,angle=0,clip]{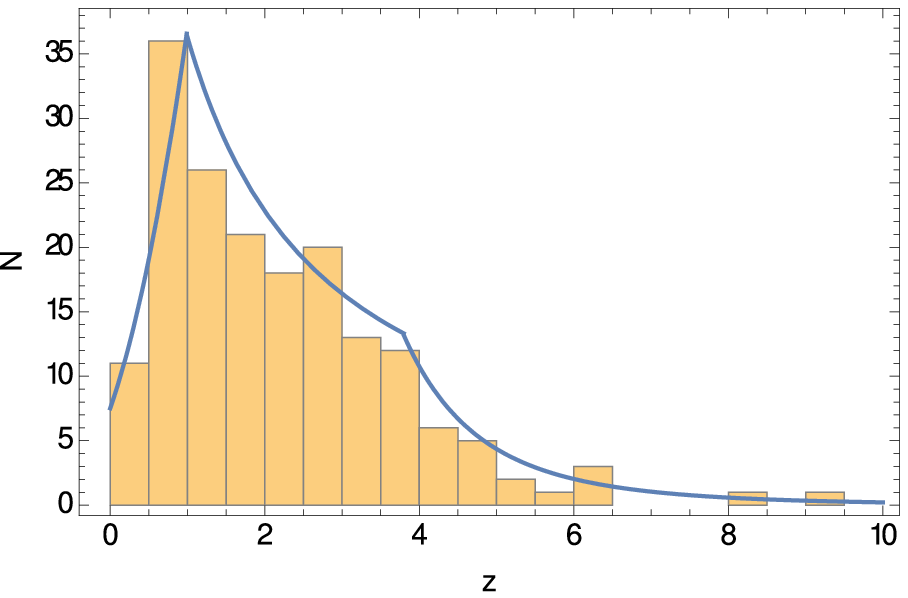}
\includegraphics[width=0.49\hsize,angle=0,clip]{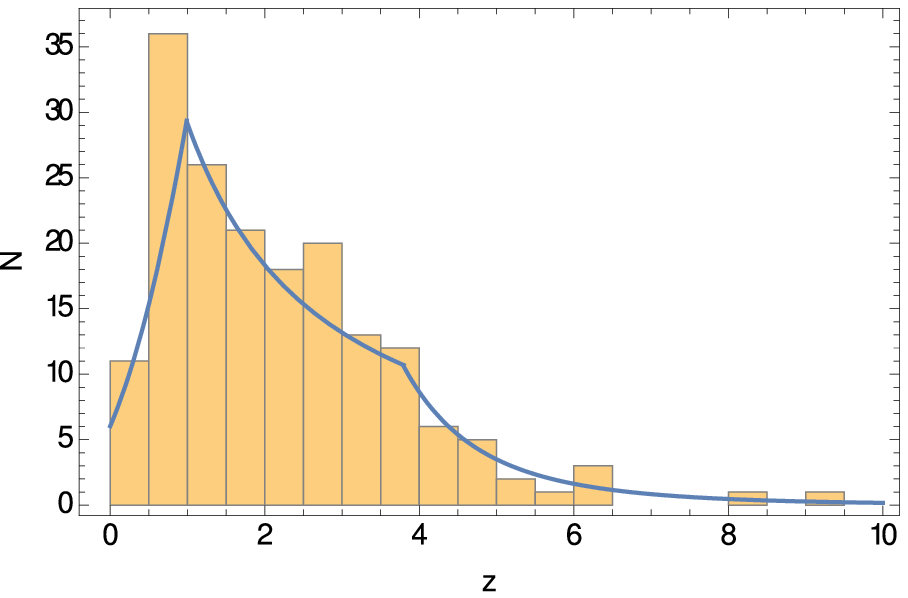}
\includegraphics[width=0.49\hsize,angle=0,clip]{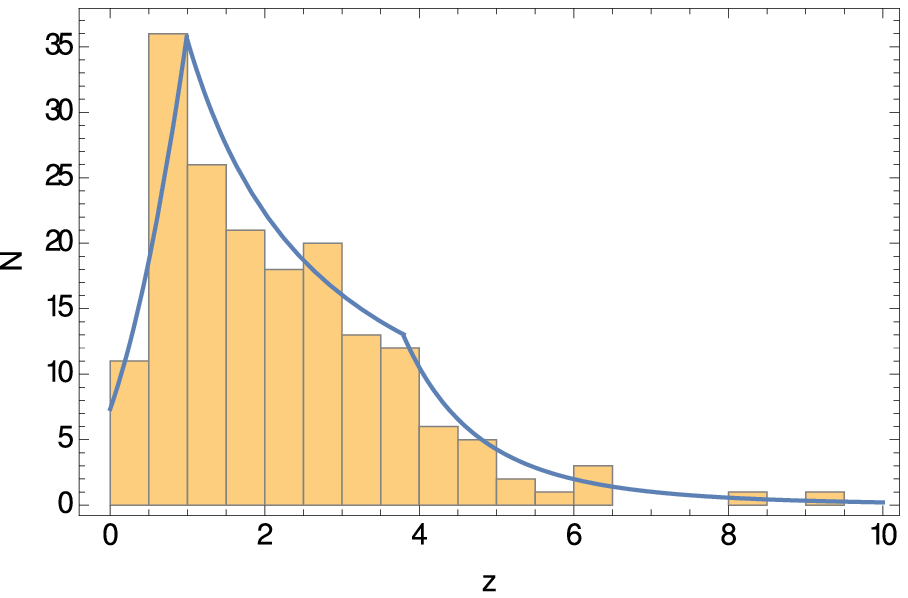}
\caption{\footnotesize GRBs rate density using method of Li (2008) and the observed GRBs rate density obtained by 
the linear efficiency functions (upper panel), and the polynomial efficiency function (lower panel) with the redshifts distribution of our data sample.}
\label{fig:3}
\end{figure}

 Nevertheless, here we study detector threshold effects on afterglow properties. Our aim is to investigate how the ignorance of the efficiency function 
bias the estimate of the correlation coefficients. We can therefore rely on simulated samples based on a realistic 
intrinsic rate. We proceed as schematically outlined below.

\begin{itemize}
\item[(i)] We assume that the available data represent reasonably 
well the intrinsic $\tau$ distribution so that we can infer ($\tau_{0}$, $\sigma_{\tau}$)
from the data themselves. We set $\tau_{L,U} = \tau_{0} \pm 5\sigma_{\tau}$
thus symmetrically cutting the Gaussian distribution at its 
extreme ends.
\item[(ii)] Based on the shape of the cosmic SFR \citep{Hopkins2006}, we assume a broken power law for the 
comoving GRB rate density:

\begin{equation}
\dot{\rho}_*(z) \propto  \begin{cases} (1+z)^{g_0} & z\le z_0 \\ (1+z)^{g_1} & z_0\le z \le z_1 \\ (1+z)^{g_2} & z \ge z_1
\end{cases}
\label{rho}
\end{equation}

 where the relative normalizations are set so that $\dot{\rho}_*(z)$ is continuous at $z_0 = 0.97$ 
and $z_1$ and  $(z_0,z_1)=(0.97,4.50)$, $(g_0 , g_1, g_2) = (3.4, -0.3, -8.0)$. Moreover, besides the equation
\ref{rho}, we employed other shapes of the SFR \citep{Li08,Robertson2012,kistler13} to obtain the observed GRBs rate density. The one used by \citep{Li08} is:

\begin{equation}
\dot{\rho}(z)=a + b\times Log(1 + z). 
 \end{equation}

The a and b parameters are :

\begin{equation}
(a, b) = \begin{cases} (-1.70, 3.30) & z \le 0.993 \\ (-0.727, 0.0549) & 0.993 \leq z \leq 3.80 \\ (2.35,-4.46) & z \geq 3.80 \\
\end{cases}
\end{equation}

Robertson \& Ellis (2012) defined the SFR as:
\begin{equation}
\dot{\rho}(z) = \frac{a+b(z/c)^f}{1+(z/c)^d}+g,
\end{equation}
where they have $a = 0.009M_{\odot} yr^{-1} Mpc^{-3}$, $b = 0.27M_{\odot} yr^{-1} Mpc^{-3}$, $c=3.7$, $d = 7.4$, and $g = 10^{-3}M_{\odot} yr^{-1} Mpc^{-3}$.

Instead, Kistler et al. (2013) defined the SFR as :

\begin{equation}
\dot{\rho}(z) = \dot{\rho}_0 \times [(1 + z)^{a \psi} + (\frac{1 + z}{B})^{b \psi} + (\frac{1 + z}{C})^{c \psi}]^{\frac{1}{\psi}},
\label{densityfunc}
\end{equation}

with slopes $a= 3.4$, $b=-0.3$, and $c=-2.5$, breaks at $z_1 = 1$ and $z_2 = 4$ corresponding to 
$B = (1+z_1)^{1-\frac{a}{b}} \sim 5160$ and $C=(1+z_1)^{\frac{(b-a)}{c}}\times(1+z_2)^{\frac{(1-b)}{c}} \sim 11.5$, 
and $\psi=-10$.

Finally, we compare the fitted functions obtained with these four methods with our
data distribution. The most realible fits for our parameters is the SFR used by Li (2008), see Fig. \ref{fig:3} where
the best fit among linear (upper panel) and polynomial (lower panel) $\epsilon(\lambda)$ functions are considered.
Moreover, we adopted the constraints for the redshift dependent ratio between SFR and GRB rate adopted by 
Robertson \& Ellis (2012). In this paper a modest evolution (e.g.,$\Psi(z)\approx(1+z)^{\alpha}$) with 
$-0.2 \leq \alpha \leq 1.5$, where the peak probability occurs for $\alpha \approx 0.5$ is consistent with the long 
GRB prompt data ($P\approx0.9$). These values can be explained if GRBs occur primarily in low-metallicity galaxies which are proportionally more numerous at earlier times. We note that in our approach we assumed no evolution at low redshift for $z \leq 0.99$ consistently with the posterior probability in Robertson \& Ellis (2012) in which no evolution is possible at the 2-$\sigma$ level. However, because a constant $\Psi(z)$ is also ruled out \citep{Robertson2012}, then we fit the normalization parameters and the evolution factors obtaining $\Psi(z)\approx(1+z)^{-0.2}$ for $0.993 \leq z \leq 3.8$ and $\Psi(z)\approx(1+z)^{0.5}$ for $z \ge 3.8$. These values of the evolution are compatible with Robertson et al. (2012). Regarding the observed GRB rate we obtained that the best efficiency functions are possible both for two polynomial and two linear as we show in Fig. 4. Table \ref{tbl1} and \ref{tbl2} show the probability that the density rate match the afterglow plateau GRB rate assuming those efficiency functions

\item[(iii)] For given ($\alpha_0$, $\alpha_{\tau}$, $\alpha_{\zeta}$, $\sigma_{int}$) values, we divide the 2D space
($\lambda$, $z$) in $\mathcal{M}$ cells and, for each cell, compute the fraction of GRBs in it as:

\begin{equation}
f_{sim}(\lambda_i,z_i)=\frac{\int_{\lambda_i-\Delta \lambda}^{\lambda_i+\Delta\lambda}d\lambda \int_{z_i-\Delta z}^{z_i+\Delta z}dz \frac{d\mathcal{N}}{d\lambda dz}}{\int_{\lambda_{min}}^{\lambda_{max}}d\lambda \int_{z_{min}}^{z_{max}}dz \frac{d\mathcal{N}}{d\lambda dz}}
\label{fsim}
\end{equation}

 where we set
\begin{equation}
(\lambda_{min}, \lambda_{max}) = (42.0, 52.0),(z_{min}, z_{max})=(0,10). 
\label{val2}
\end{equation}

 We find more efficient to change variable from $z$ to $\zeta$ when
dividing the 2D space in $10 \times 10$ square cells.

\item[(iv)] For each given cell, we generate $\mathcal{N}_{ij}=f_{sim}(\lambda_i, \zeta_j) \times \mathcal{N}_{sim}$ GRBs (with $\mathcal{N}_{sim}$
the total number of objects to simulate) by randomly sampling ($\lambda$, $\zeta$) within the cell boundaries and 
computing $\tau$ by solving Eq. \ref{scaling}.
\item[(v)] To take into account of the selection effects, for each GRB, we generate two random numbers 
($u_{\tau}$,$u_{\lambda}$) uniformly sampling the range (0, 1) and only retain the GRB if $u_{\tau}\le\mathcal{E}_{\tau}(\tau)$ 
and $u_{\lambda}\le\mathcal{E}_{\lambda}(\lambda)$. Note that, as a consequence of
this cut, the final number $\mathcal{N}_{obs}$ of observed GRBs is smaller than the input one $\mathcal{N}_{sim}$.
\item[(vi)] Finally, for each one of the $\mathcal{N}_{obs}$ selected GRBs, we generate new ($\tau_{obs}$, $\lambda_{obs}$) 
values extracting from Gaussian distributions centered on the simulated ($\tau$, $\lambda$) values and with a 1\% 
variance. We also associate to each GRB an error set in such a way to be similar to what is actually obtained for GRBs 
having comparable ($\tau$, $\lambda$) values.
\end{itemize}

 The above procedure allows us to build simulated GRBs
sample taking into account both the intrinsic properties of
any scaling relation and the selection effects induced by the
instrumental setup. Moreover, we have referred to an actual
GRBs sample in order to set both the limits on ($\tau$, $\zeta$, $\lambda$)
and the typical measurement errors. Therefore, we can  rely
on these simulated samples to investigate the impact of
selection effects on the recovered slope and intrinsic scatter
of the given correlation. To this end, the last ingredient we
need is a functional expression for the efficiency functions.
Since these are largely unknown, we are forced to make
some arbitrary guess. Therefore, we consider two different
cases. First, we assume that there is no selection on $\tau$ , i.e. we set
$\mathcal{E}_{\tau} = 1$. Two functional expressions are then used for $\mathcal{E}_{\lambda}$,
namely a power law:

\begin{equation}
\mathcal{E}_{\lambda}(\lambda) = \begin{cases} 0 & \lambda<\lambda_L \\ (\frac{\lambda-\lambda_L}{\lambda_U-\lambda_L})^{\mathcal{E}_{\lambda}} & \lambda_L\le \lambda \le \lambda_U \\ 1 & \lambda>\lambda_U
\end{cases} 
\label{E1}
\end{equation}

 and a fourth order polynomial, i.e. :

\begin{equation}
\mathcal{E}_{\lambda}(\lambda) = \begin{cases} 0 & \lambda<\lambda_L \\ \frac{\mathcal{E}_1\tilde{\lambda}+\mathcal{E}_2\tilde{\lambda}^2+\mathcal{E}_3\tilde{\lambda}^3+\mathcal{E}_4\tilde{\lambda}^4}{\mathcal{E}_1+\mathcal{E}_2+\mathcal{E}_3+\mathcal{E}_4} & \lambda_L\le \lambda \le \lambda_U \\ 1 & \lambda>\lambda_U
\end{cases}
\label{E2}
\end{equation}

\begin{figure}[htbp]
\centering
\includegraphics[width=0.42\hsize,height=0.225\hsize,angle=0,clip]{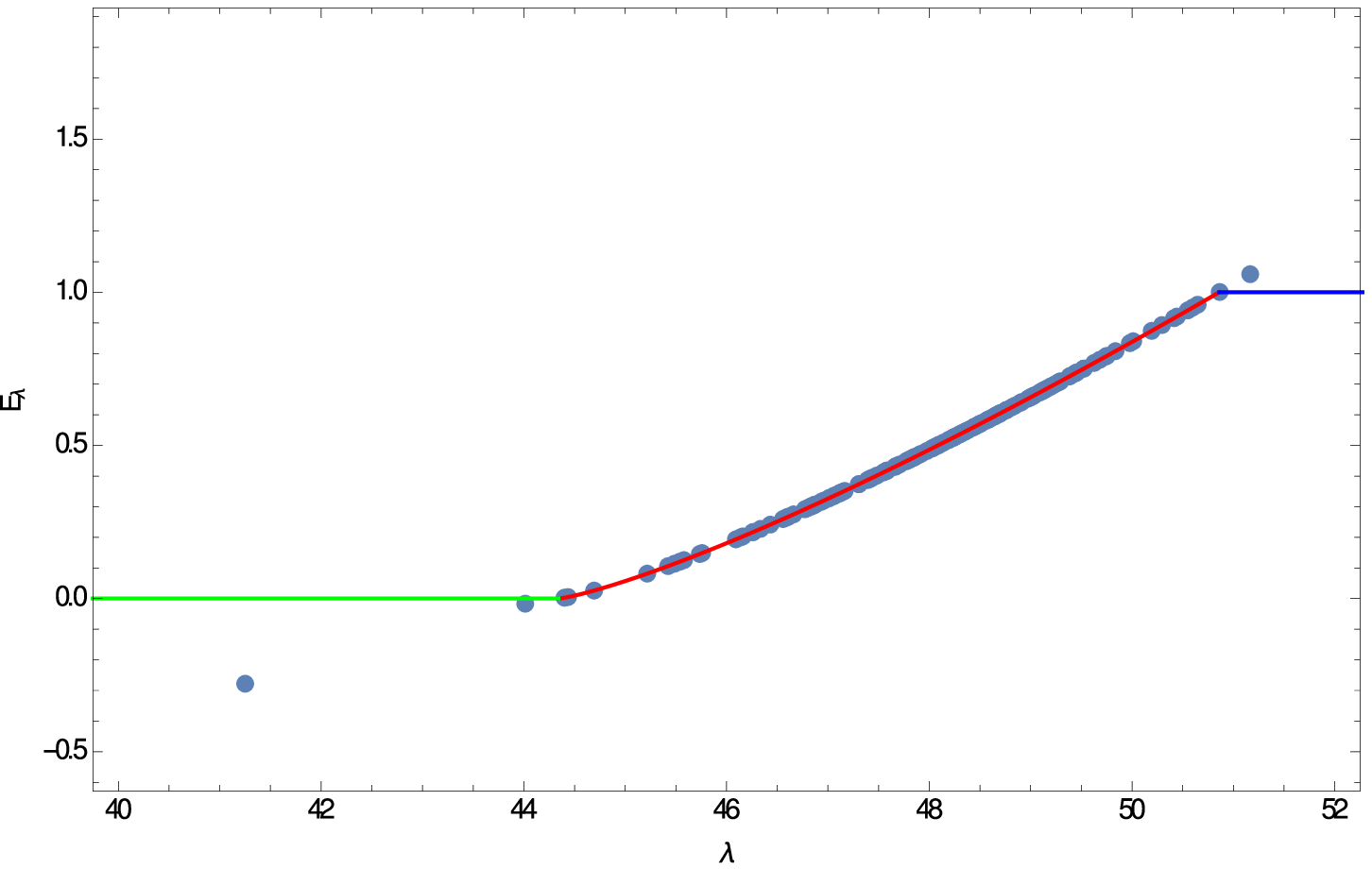}
\includegraphics[width=0.42\hsize,height=0.225\hsize,angle=0,clip]{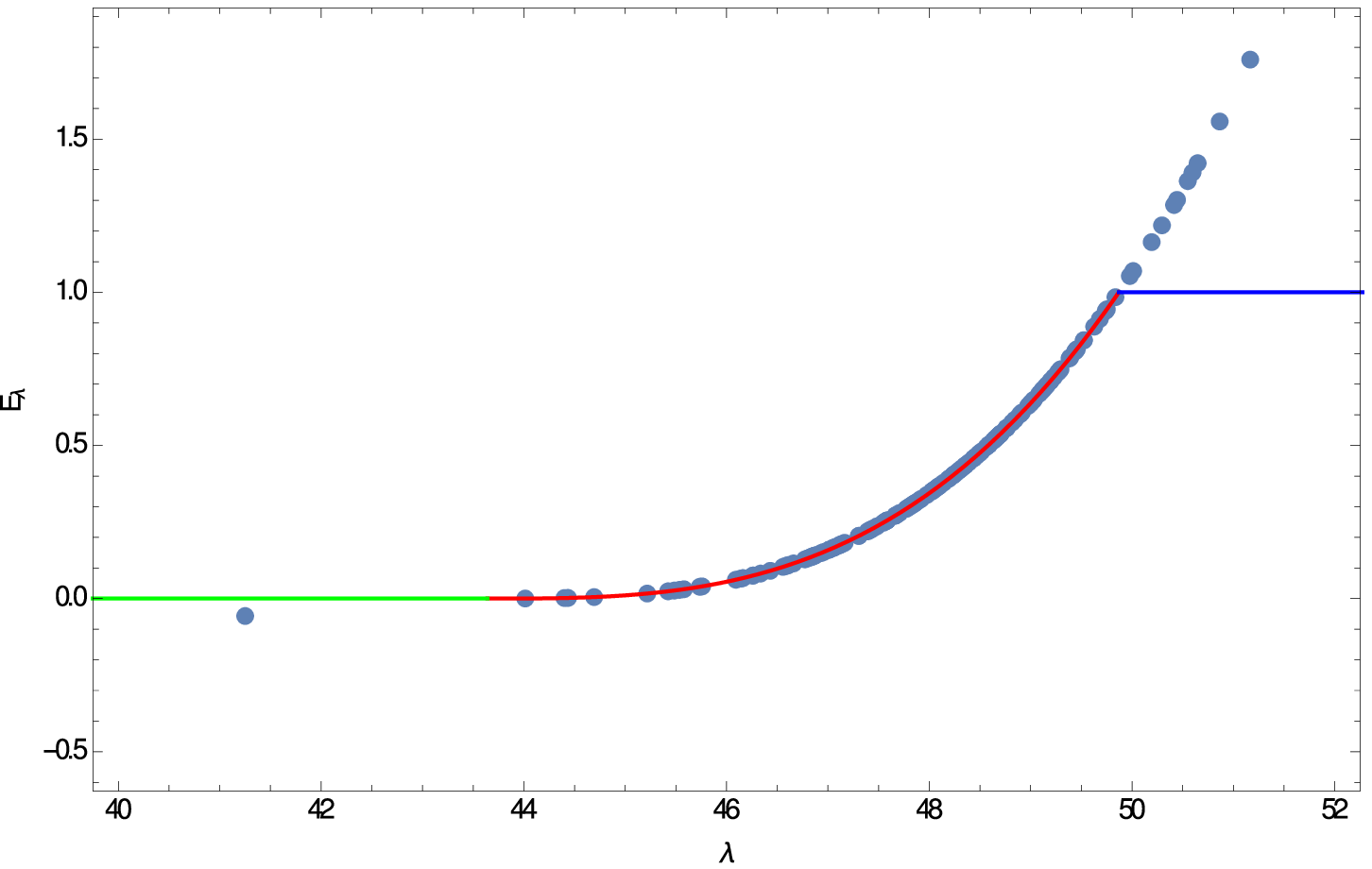}
\includegraphics[width=0.42\hsize,height=0.225\hsize,angle=0,clip]{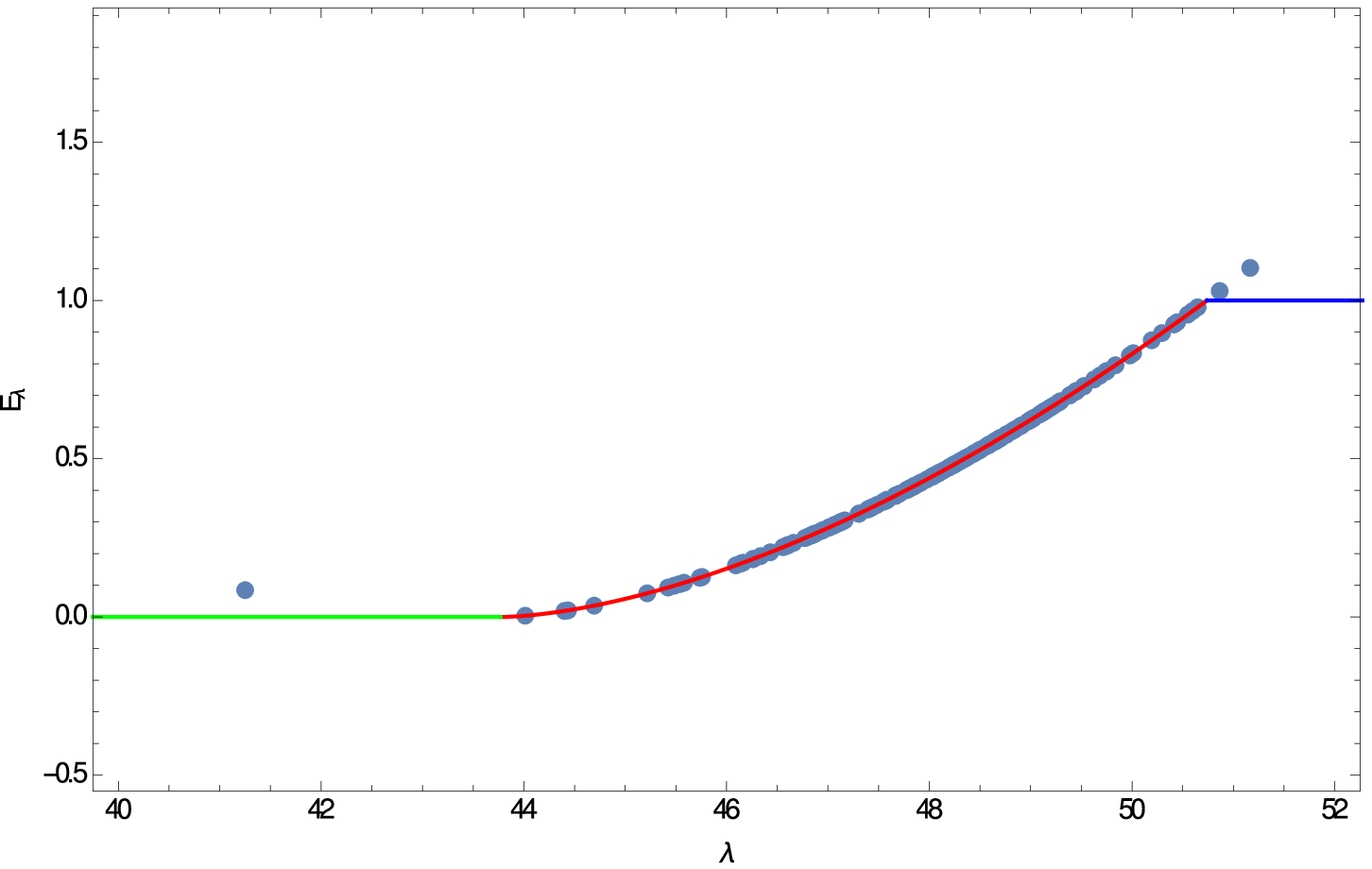}
\includegraphics[width=0.42\hsize,height=0.225\hsize,angle=0,clip]{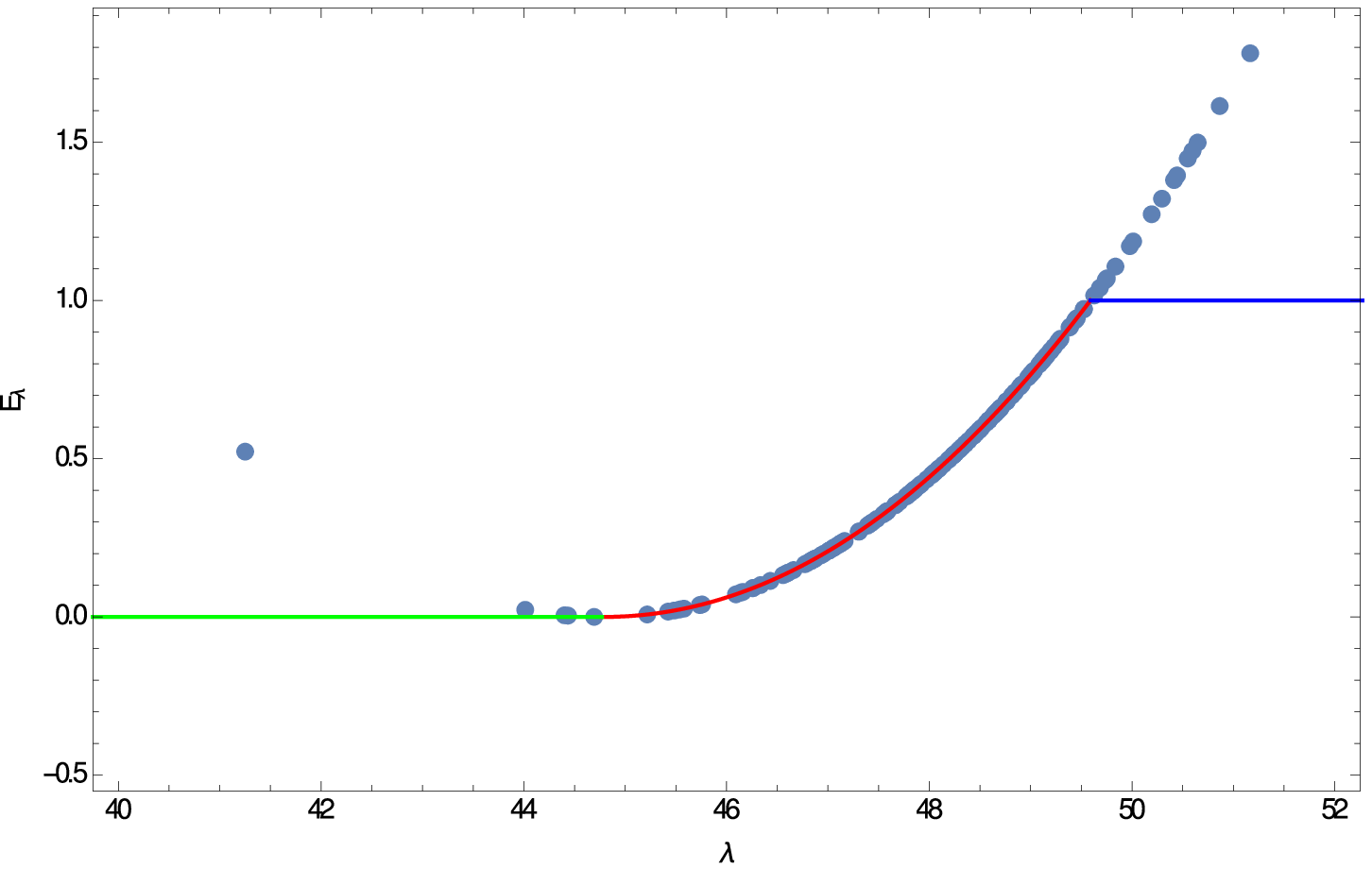}
\includegraphics[width=0.42\hsize,height=0.225\hsize,angle=0,clip]{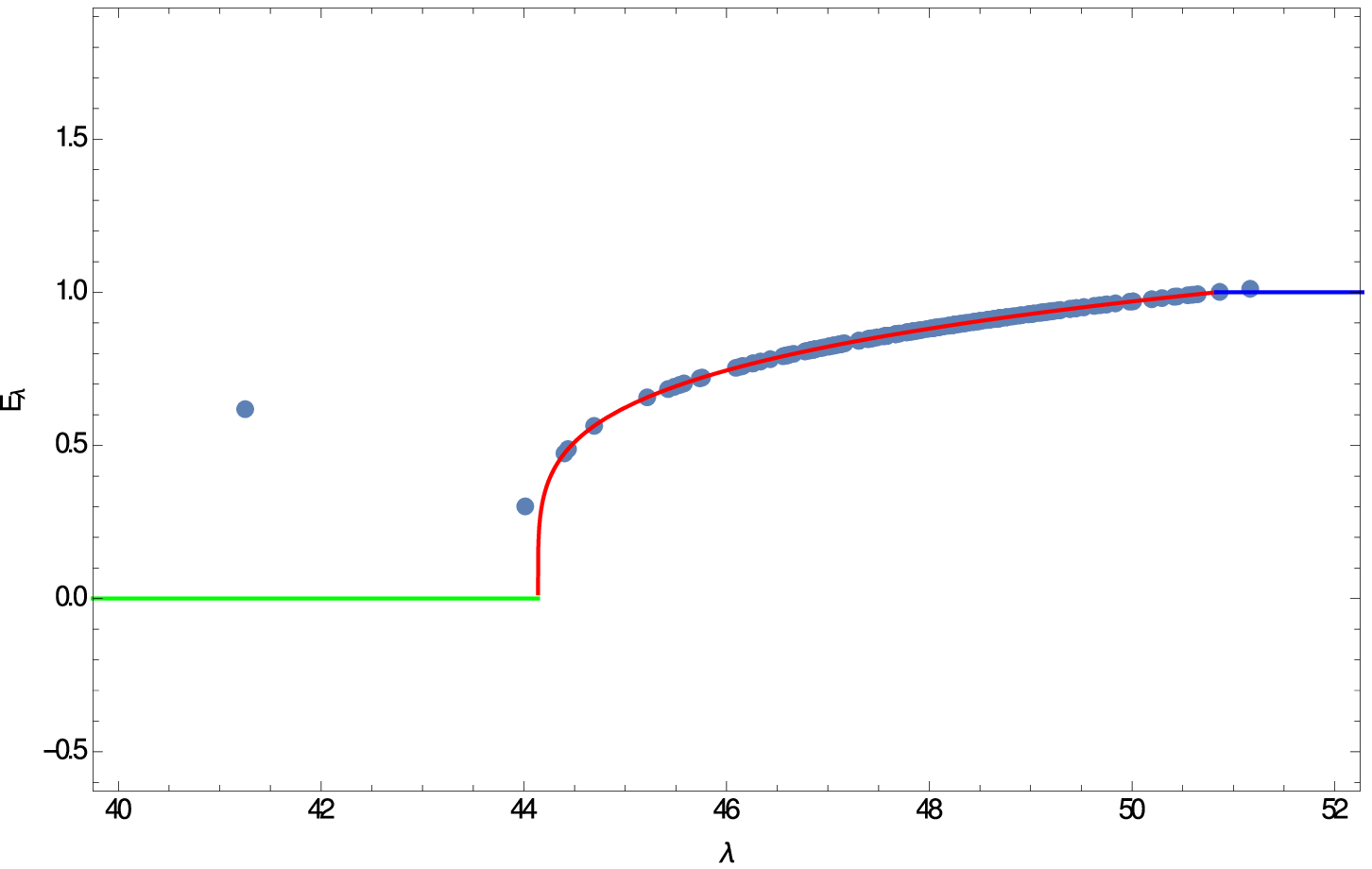}
\includegraphics[width=0.42\hsize,height=0.225\hsize,angle=0,clip]{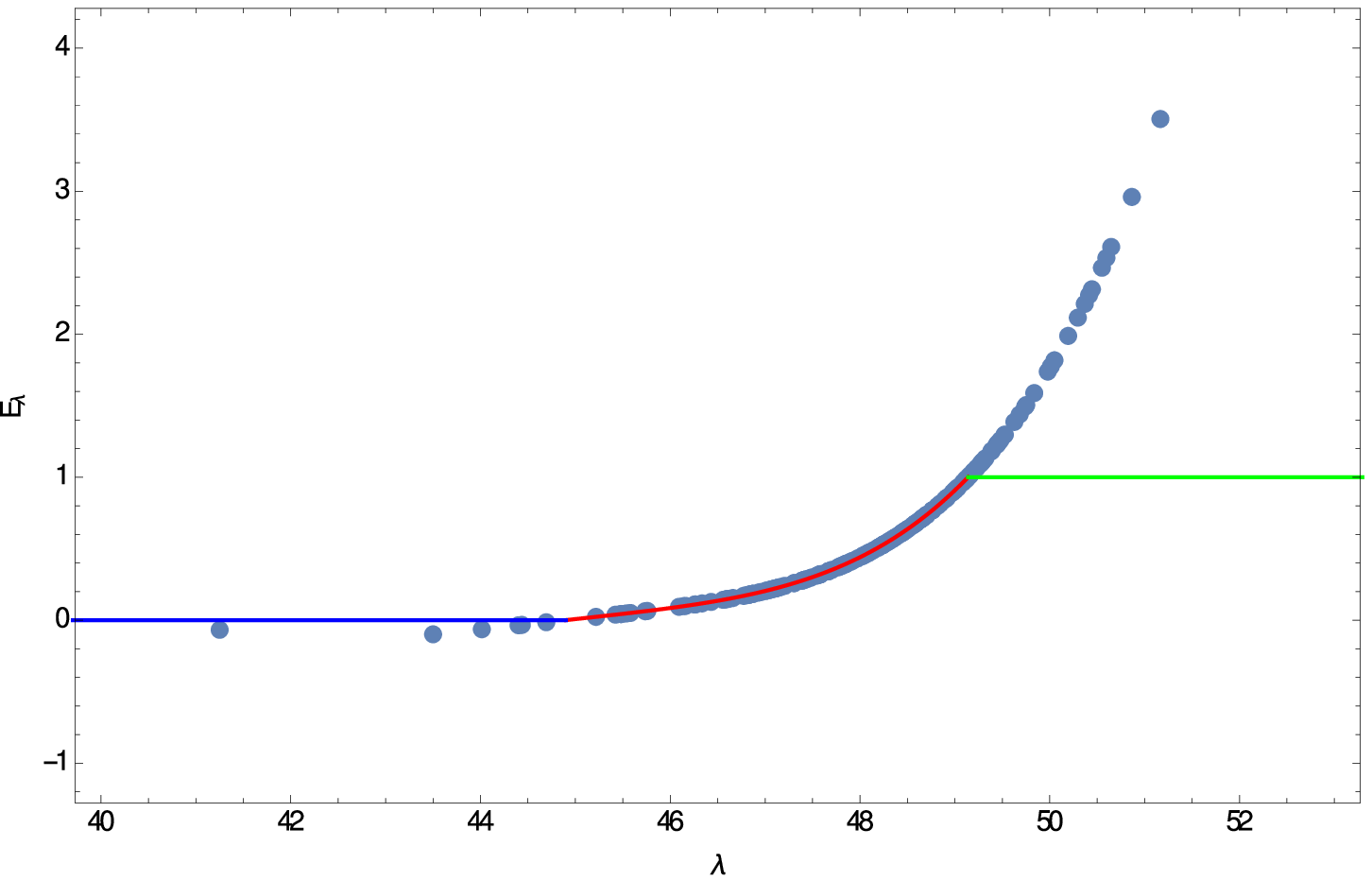}
\includegraphics[width=0.42\hsize,height=0.225\hsize,angle=0,clip]{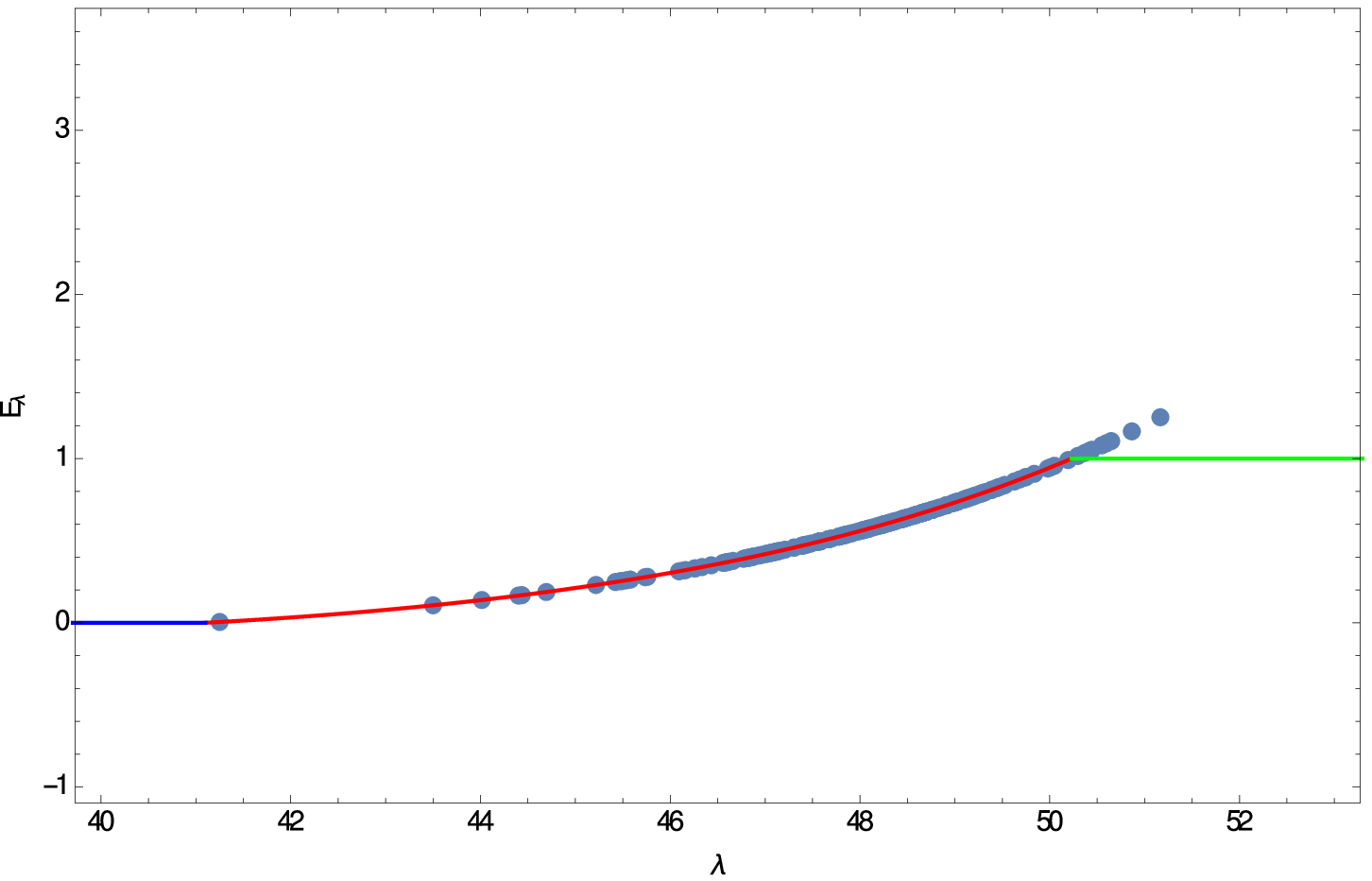}
\includegraphics[width=0.42\hsize,height=0.225\hsize,angle=0,clip]{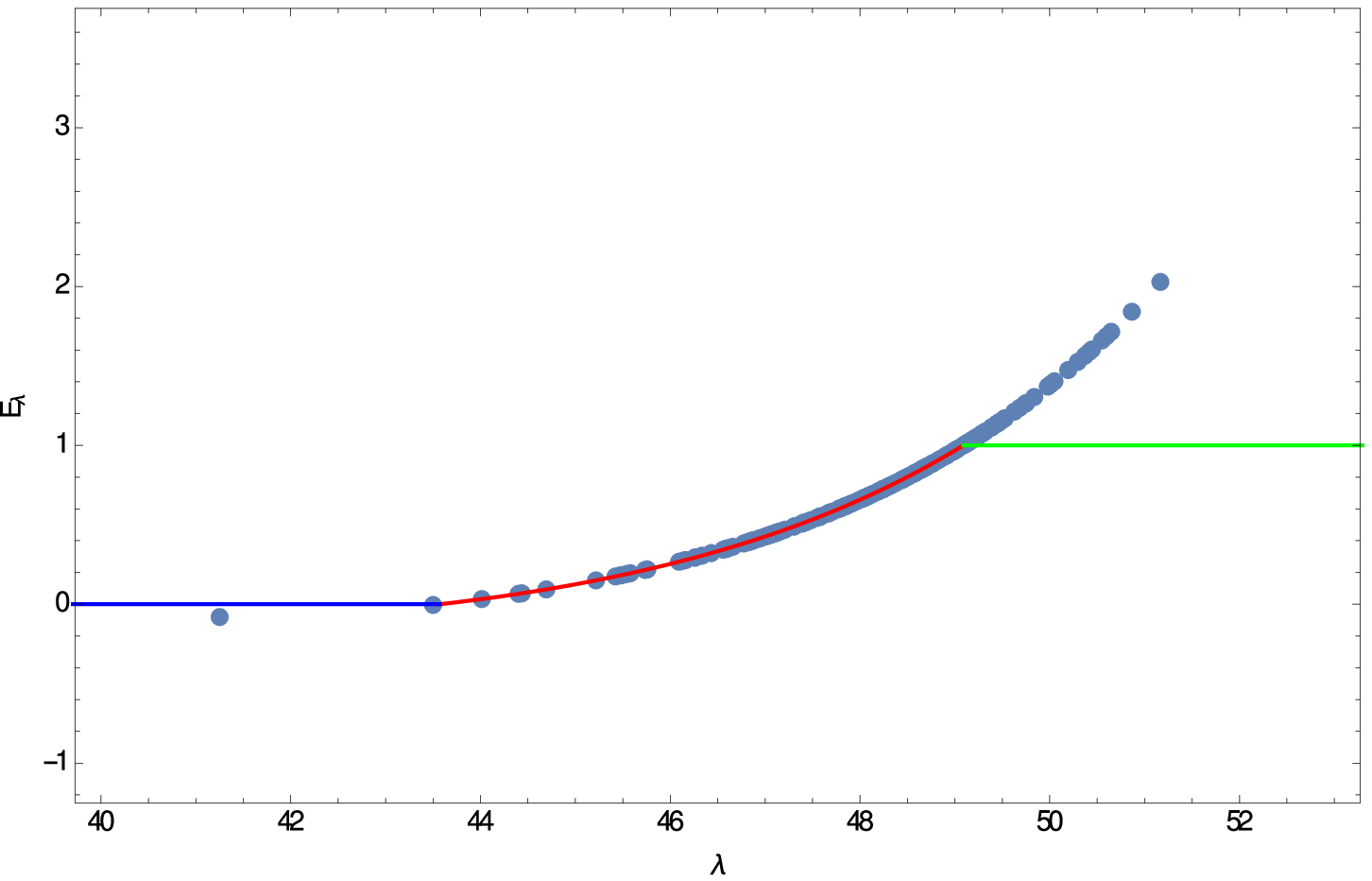}
\includegraphics[width=0.42\hsize,height=0.225\hsize,angle=0,clip]{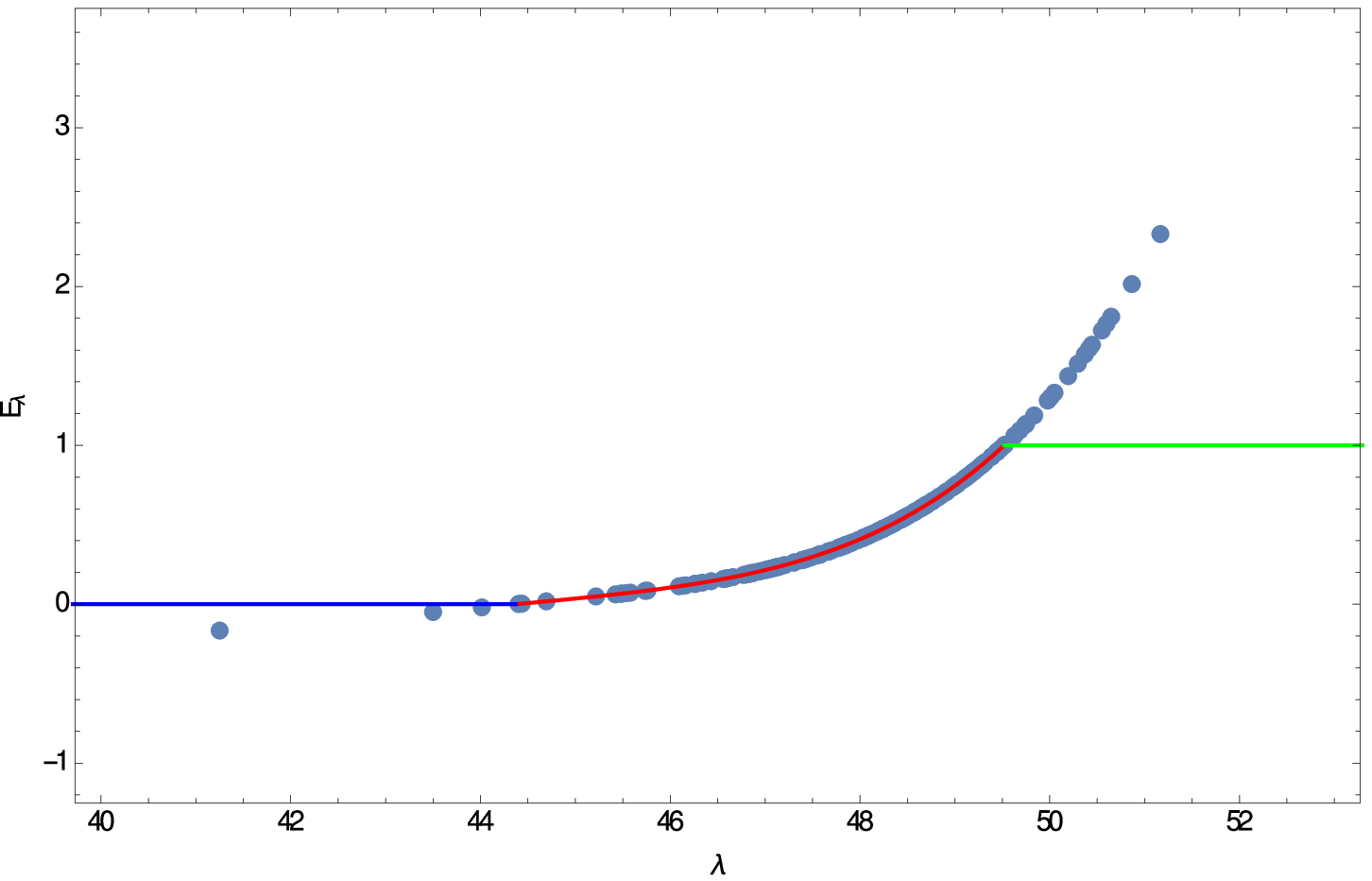}
\includegraphics[width=0.42\hsize,height=0.225\hsize,angle=0,clip]{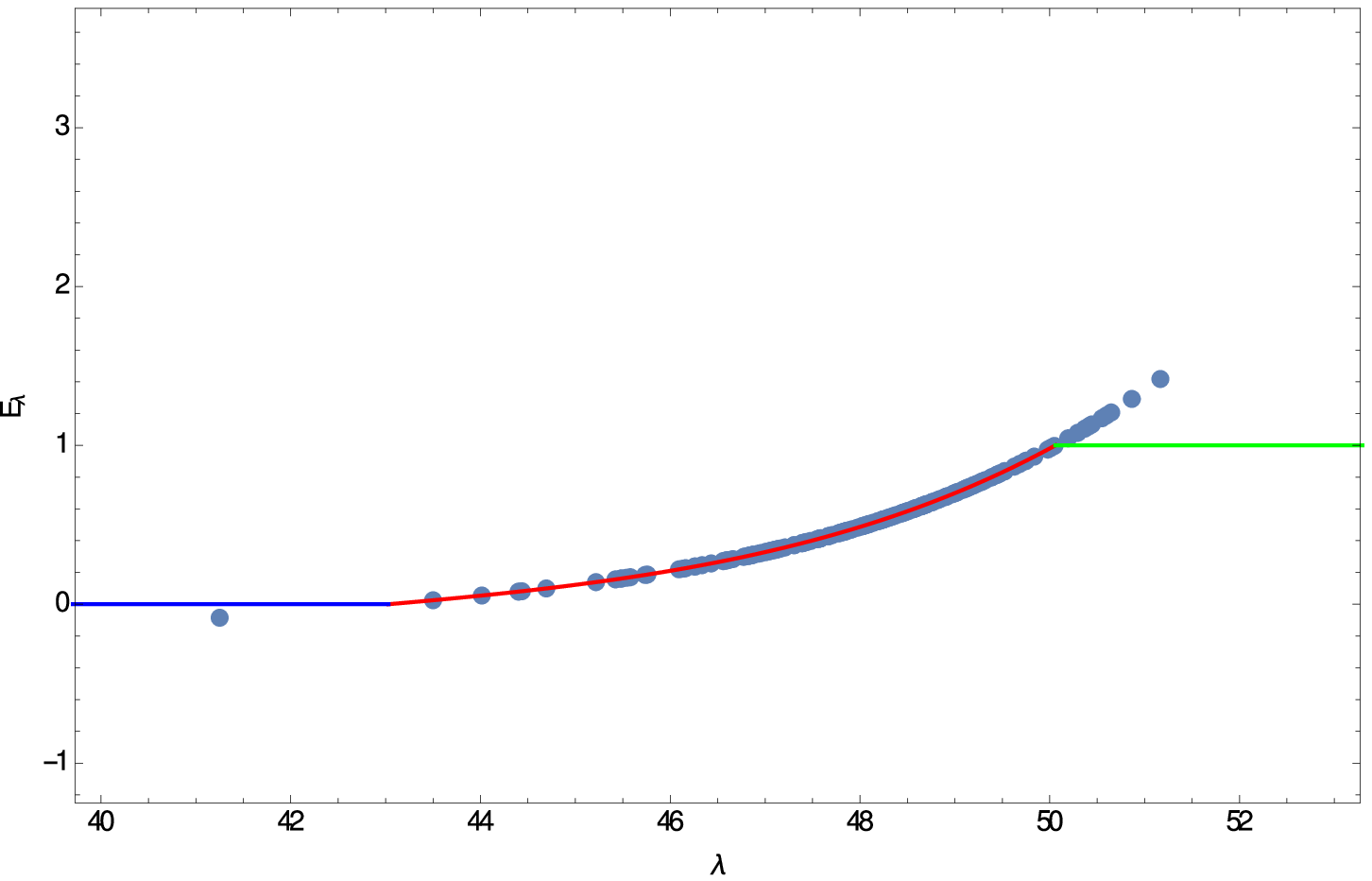}
\caption{\footnotesize The first 5 panels represent examples of the efficiency function for the linear case versus 
luminosities of the GRBs, $\lambda$, in our data sample, while the last 5 panels the efficiency functions for the 
fourth order polynomial. The linear functions as well the polynomial ones are computed according to Eq. \ref{E1}, and 
Eq. \ref{E2}.}
   \label{fig:11}
\end{figure}

\begin{figure}[htbp]
\centering
\includegraphics[width=0.42\hsize,height=0.225\hsize,angle=0,clip]{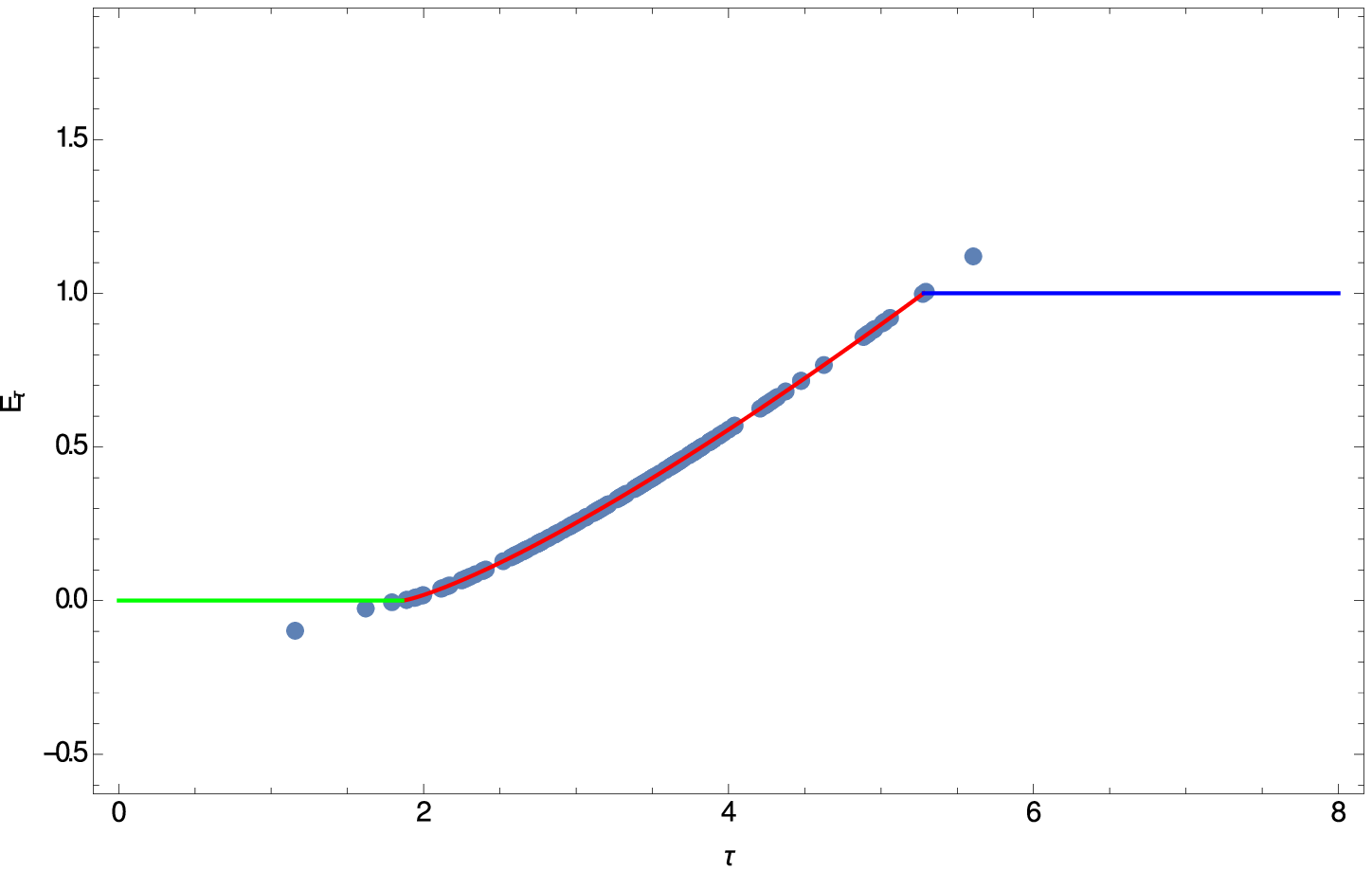}
\includegraphics[width=0.42\hsize,height=0.225\hsize,angle=0,clip]{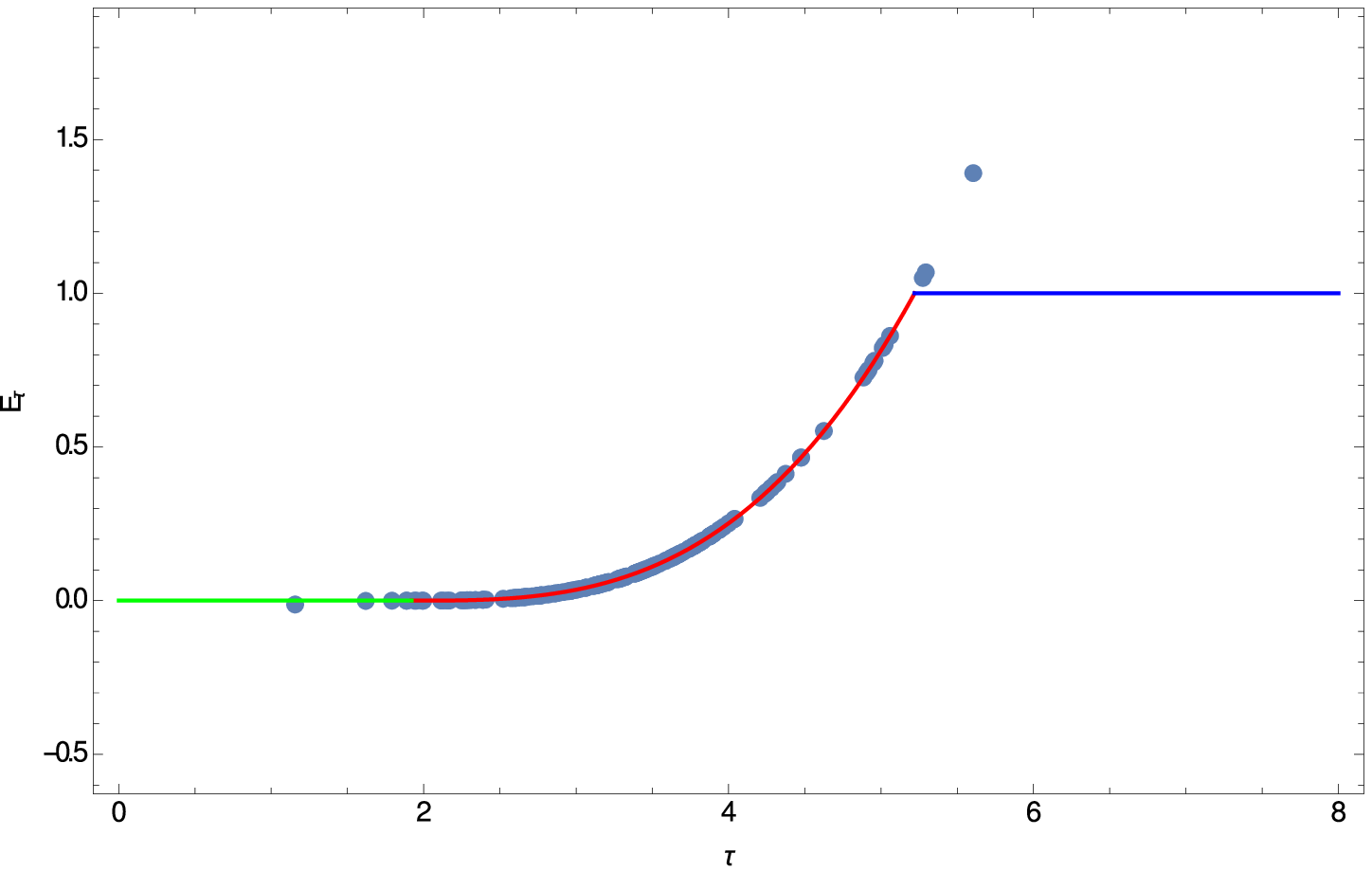}
\includegraphics[width=0.42\hsize,height=0.225\hsize,angle=0,clip]{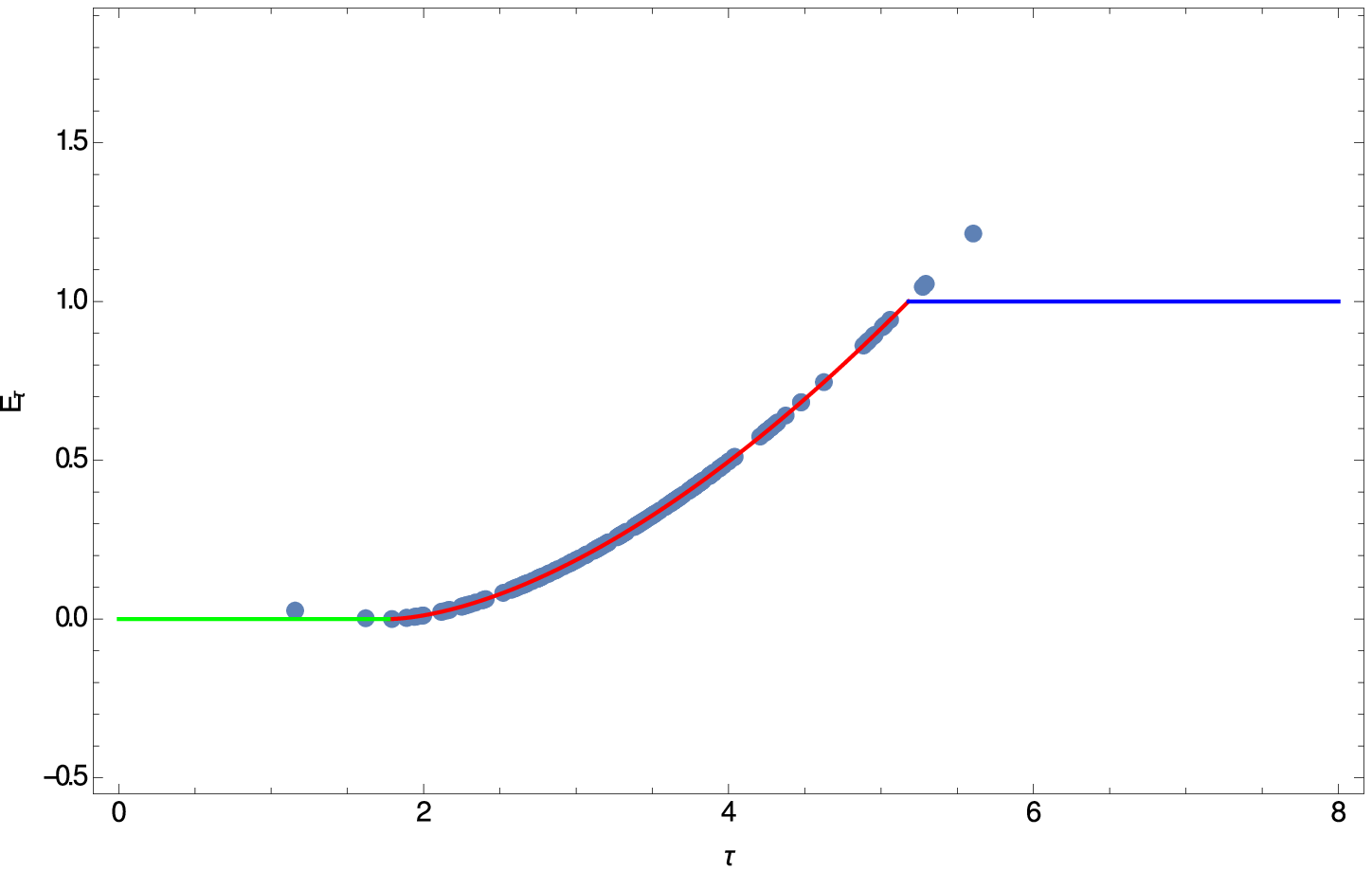}
\includegraphics[width=0.42\hsize,height=0.225\hsize,angle=0,clip]{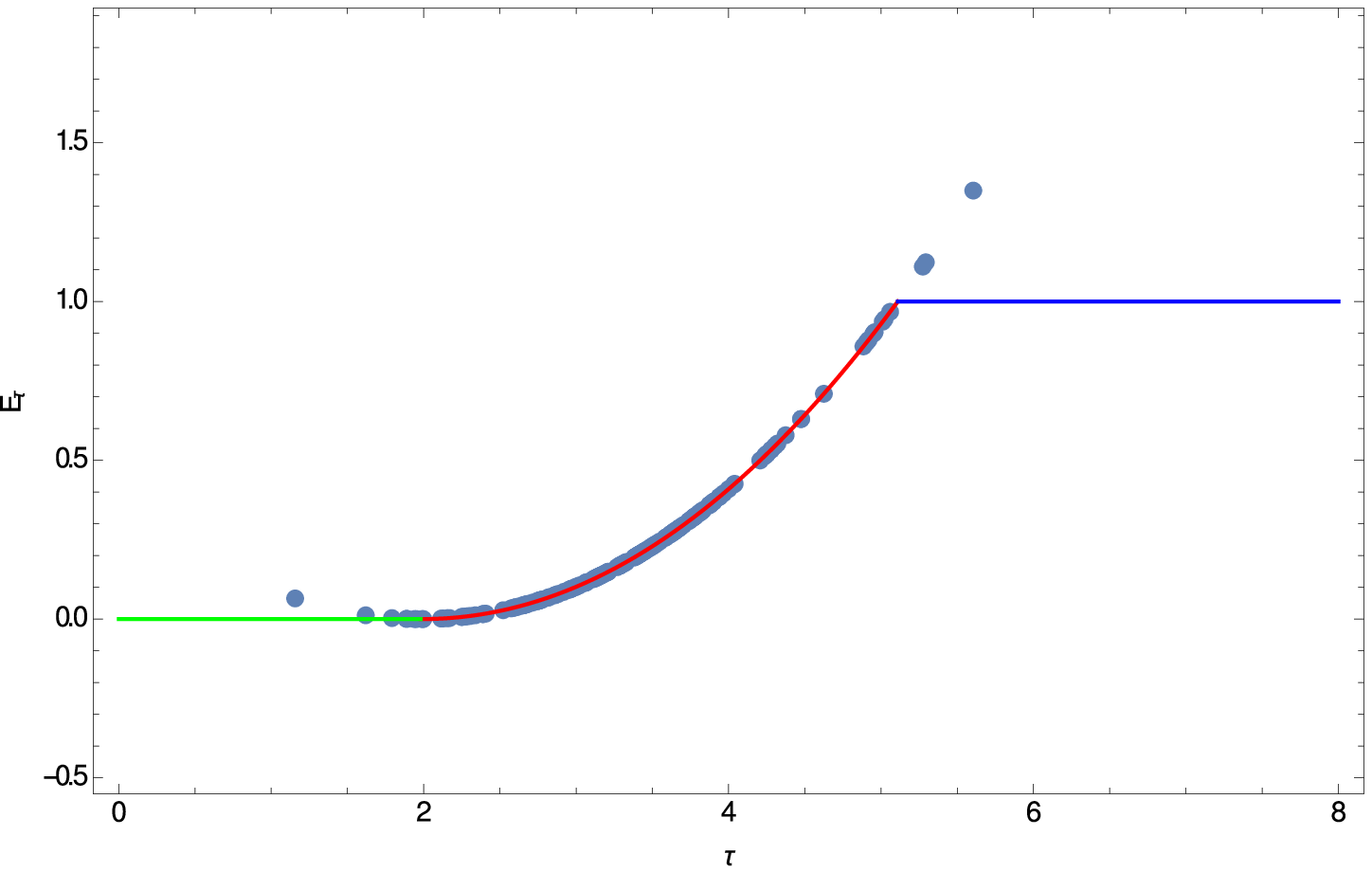}
\includegraphics[width=0.42\hsize,height=0.225\hsize,angle=0,clip]{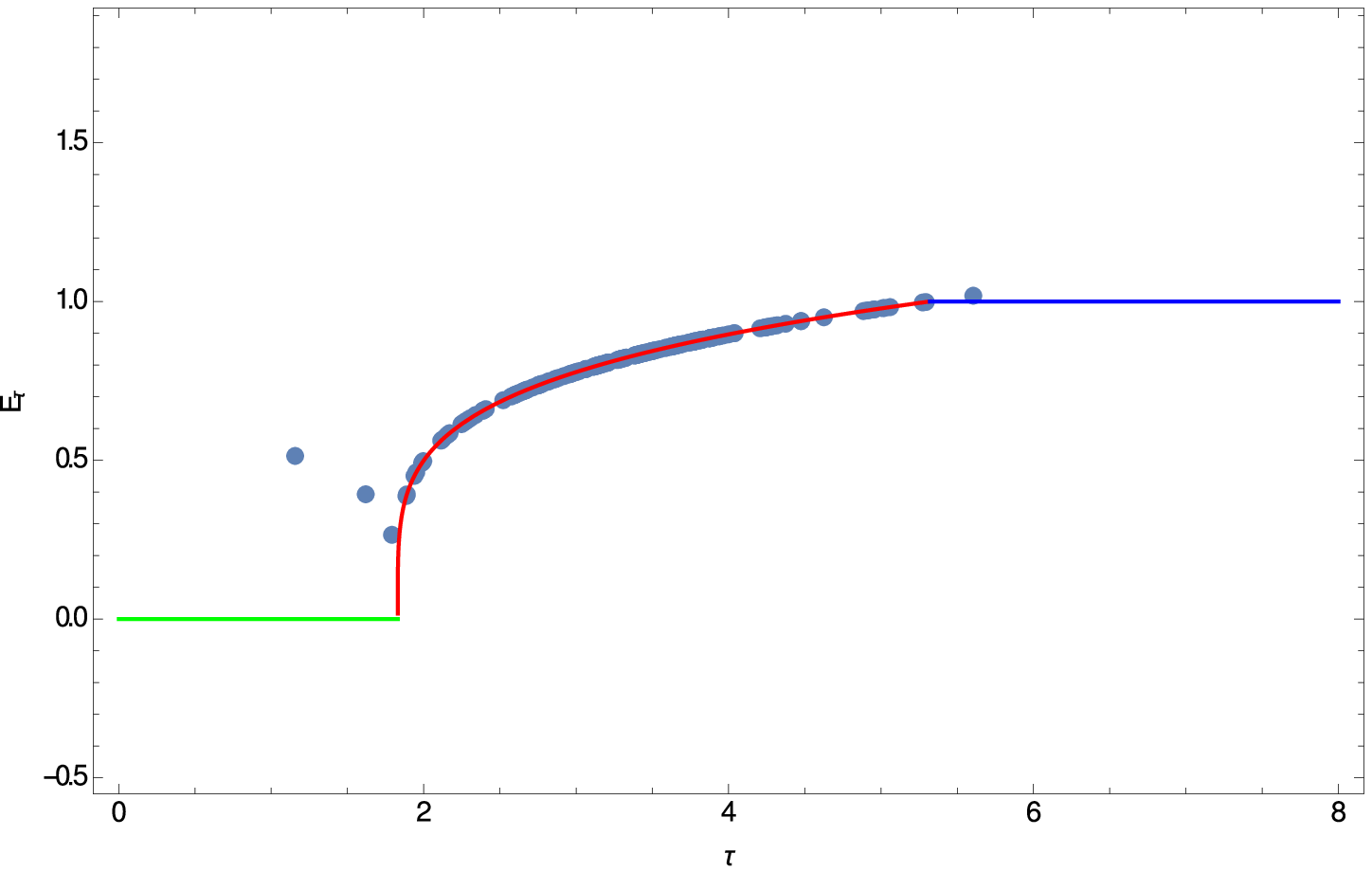}
\includegraphics[width=0.42\hsize,height=0.225\hsize,angle=0,clip]{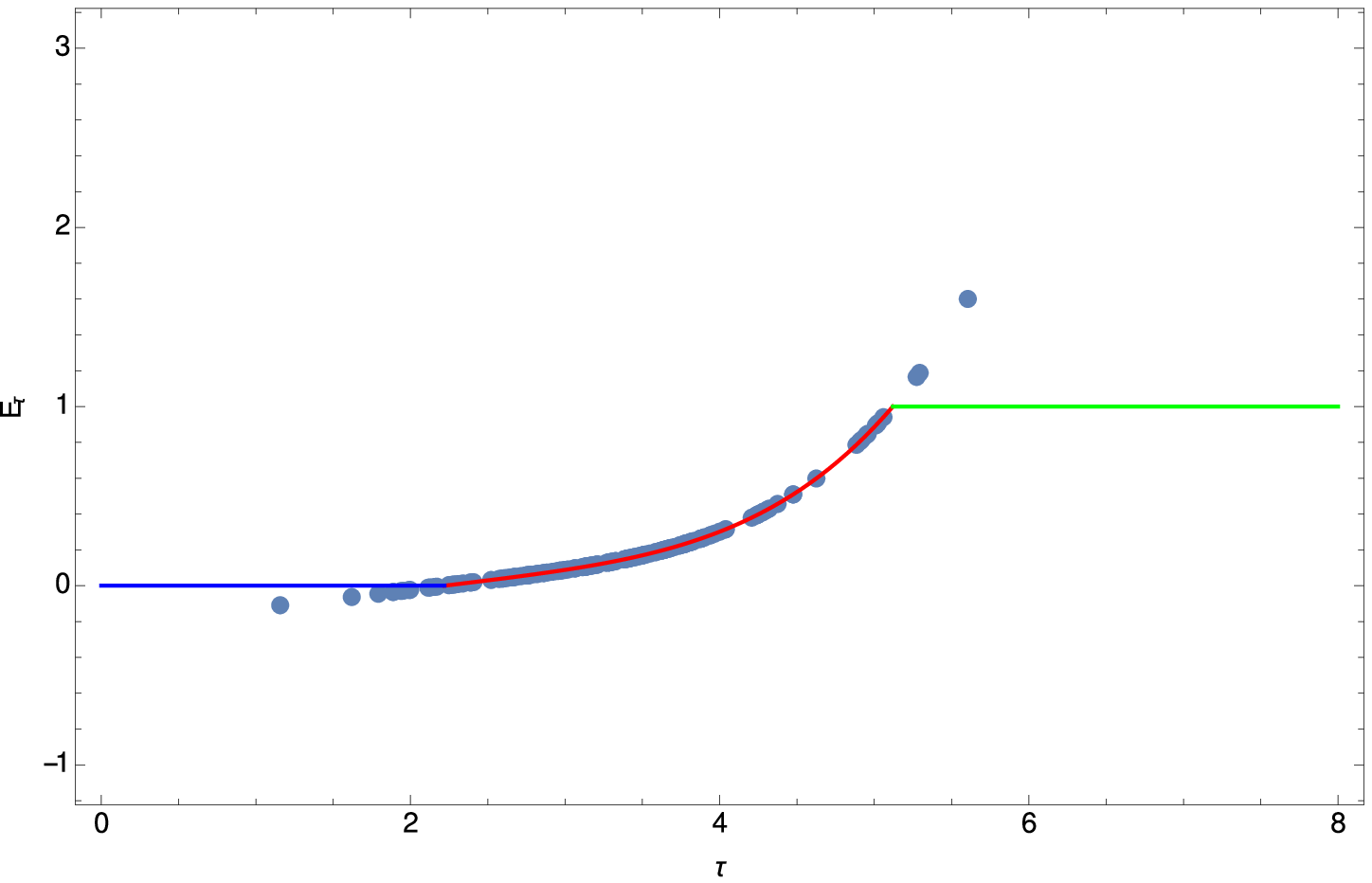}
\includegraphics[width=0.42\hsize,height=0.225\hsize,angle=0,clip]{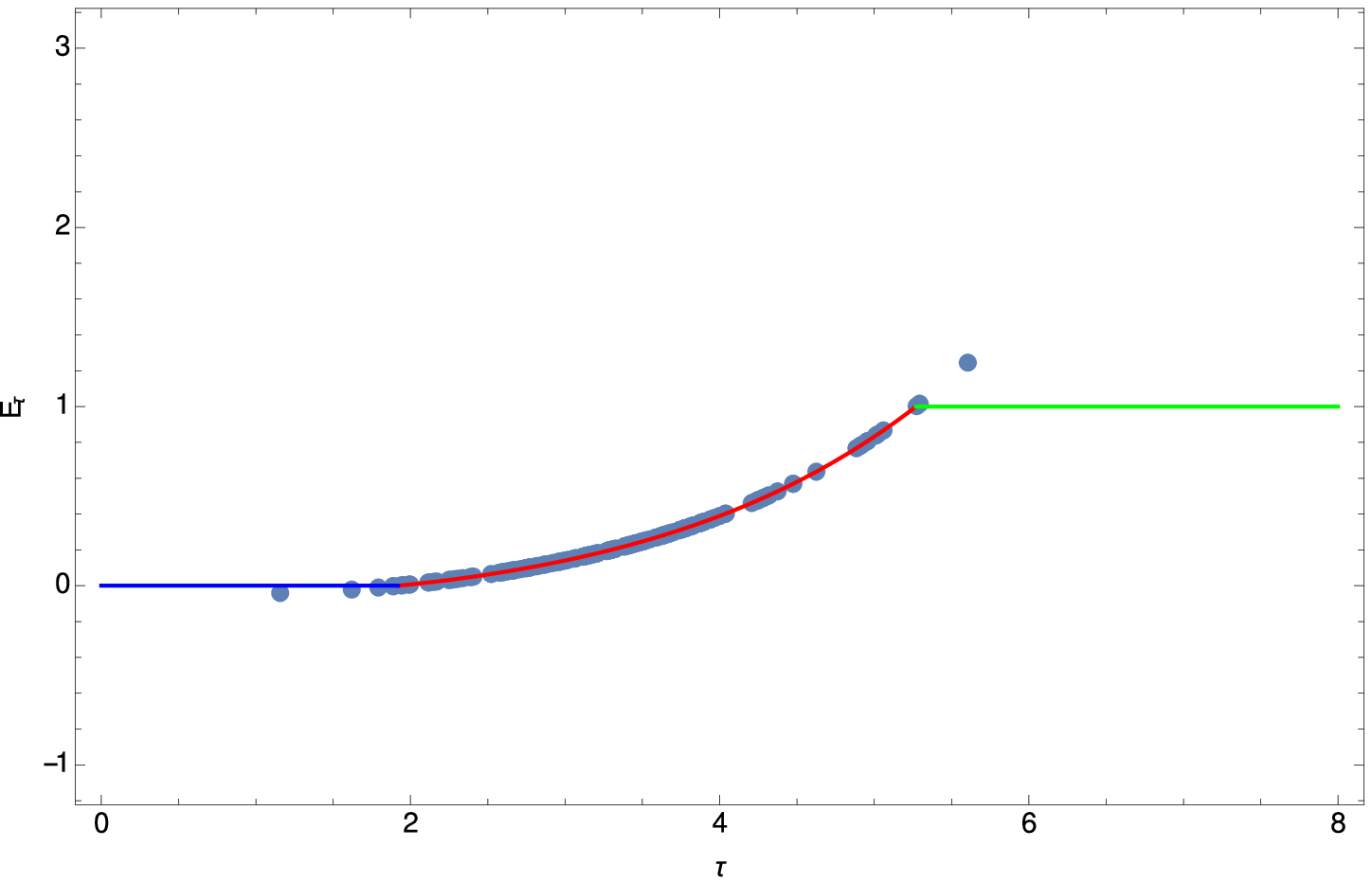}
\includegraphics[width=0.42\hsize,height=0.225\hsize,angle=0,clip]{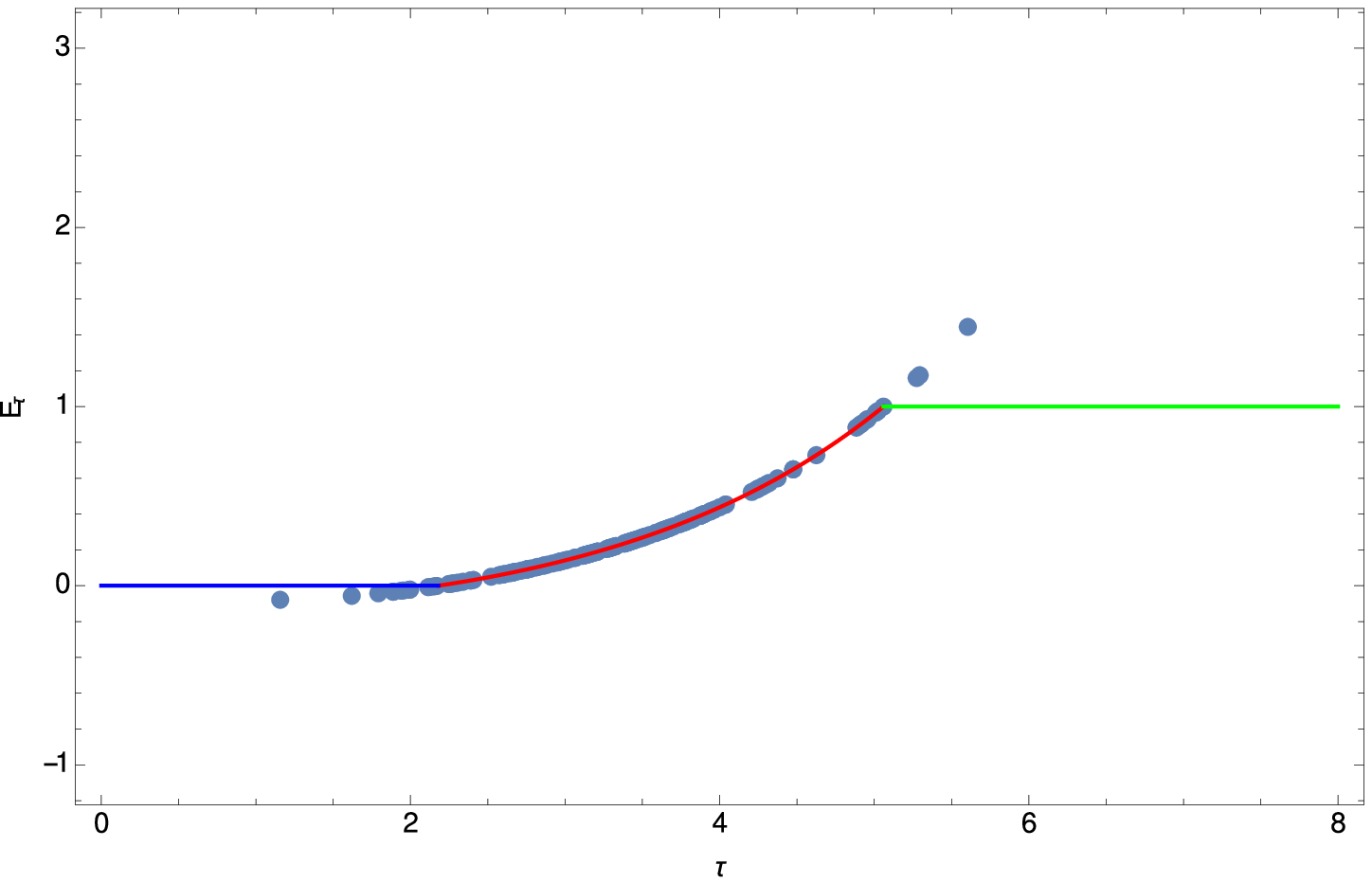}
\includegraphics[width=0.42\hsize,height=0.225\hsize,angle=0,clip]{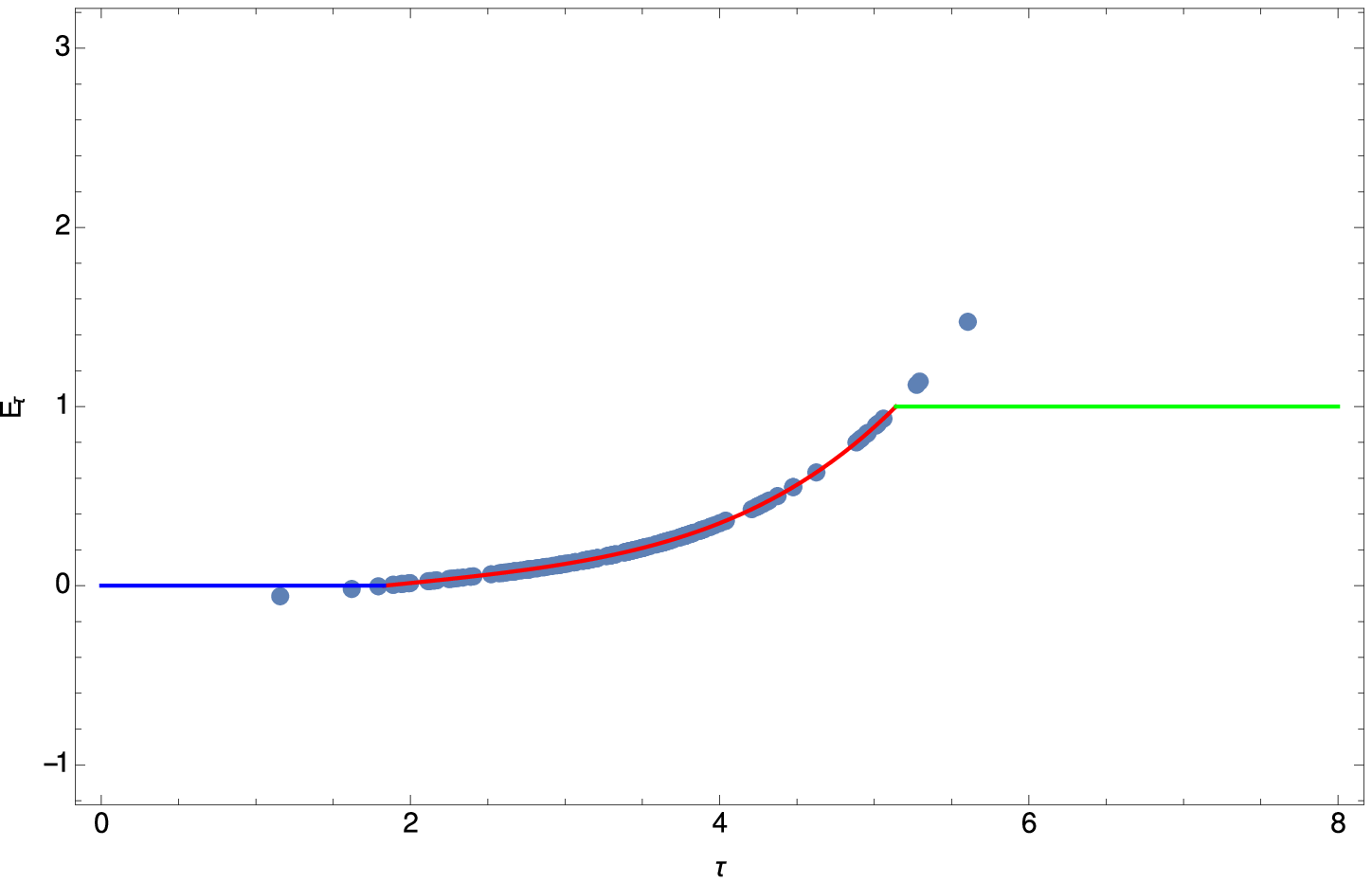}
\includegraphics[width=0.42\hsize,height=0.225\hsize,angle=0,clip]{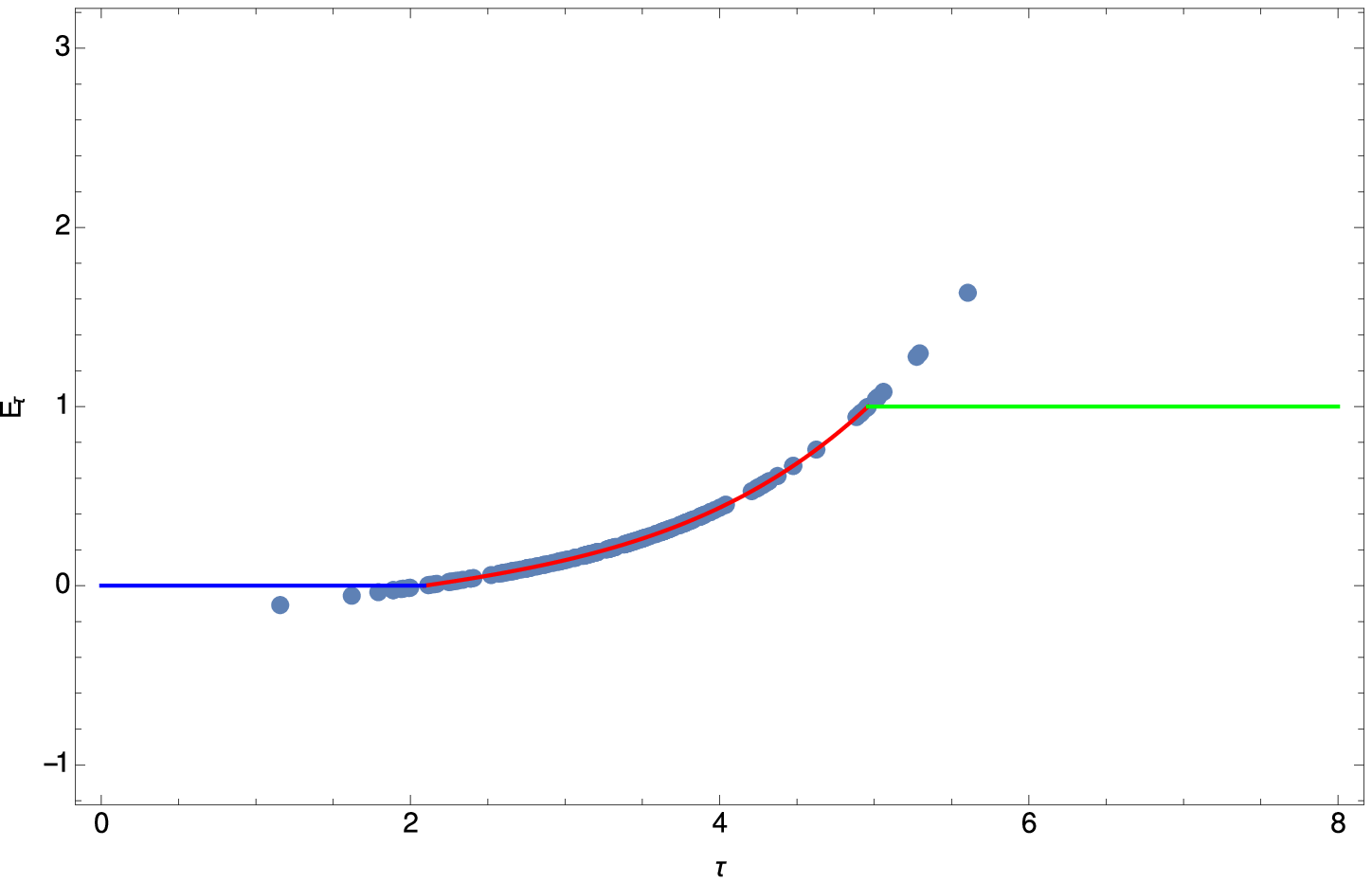}
\caption{\footnotesize The first 5 panels represent examples of the efficiency function for the linear case versus the times, $\tau$ of 
the GRBs in our data sample, while the last 5 panels the efficiency functions for the fourth order polynomial. The 
linear functions as well the polynomial ones are computed according to Eq. \ref{E1}, and Eq. \ref{E2}.}
   \label{fig:12}
\end{figure}

with $\tilde{\lambda}=(\lambda-\lambda_L)/(\lambda_U-\lambda_L)$. We try different arbitrary
choices for the parameters entering both expressions of $\mathcal{E}_{\lambda}$
in order to investigate to which extent the results depend
on the exact choice of the efficiency function, see Fig. \ref{fig:11}. In a second step, we abandon the assumption $\mathcal{E}_{\tau} = 1$,
to assume for it the same functional expression used
for $\mathcal{E}_{\lambda}$, with the same choices for the parameters, but different upper and lower limits depending on $\tau_U$ and $\tau_L$, see Fig. \ref{fig:12}.\\

\begin{table}
\footnotesize
\begin{center}{
\begin{tabular}{cccccc}
\hline
Id & $\lambda_L$ & $\lambda_U$ & $\mathcal{E}_{\lambda}$ & $P_{GRB,rate}$ \\
\hline
PL1& 44.34 & 50.86 & 1.25 & $\le 10^{-4}$ \\
PL2& 43.64 & 49.87 & 2.99 &  0.003 \\
PL3& 43.77 & 50.74 & 1.65 &  0.53 \\
PL4& 44.77 & 49.59 & 2.04 &  0.001 \\
PL5& 44.14 & 50.83 & 0.23 &  0.54\\
\hline
\end{tabular}}
\caption{Efficiency function parameters for the power-law $\mathcal{E}_{\lambda}$ and no cut on $\tau$
i.e. $\mathcal{E}_{\tau}=1$. $P_{GRB,rate}$ is the goodness of fit between our data and the observed GRBs rate density, thus how the data well fit the 
observed GRBs density rate.
To compute the probability we compute the $\chi^2$ test that performs a statistical hypothesis test in which
the sampling distribution of the test statistic is a $\chi^2$ distribution when the null hypothesis is true, in order
to determine whether there is a significant difference between the expected frequencies and the observed frequencies.}
\label{tbl1}
\end{center}
\end{table}     

\begin{table}
\footnotesize
\begin{center}
\begin{tabular}{ccccccccc}
\hline
Id & $\lambda_L$ & $\lambda_U$ & $\mathcal{E}_1$ & $\mathcal{E}_2$ & $\mathcal{E}_3$ & $\mathcal{E}_4$ & $P_{GRB,rate}$ \\
\hline
PoL1& 44.90 & 49.14 & 0.46 & 0.01 & 0.24 & 0.80 & 0.54 \\
PoL2& 41.10 & 50.23 & 0.60 & 0.95 & 0.05 & 0.53 & $\le 10^{-4}$ \\
PoL3& 43.57 & 49.09 & 0.71 & 0.79 & 0.07 & 0.34 &  0.019 \\
PoL4& 44.37 & 49.52 & 0.51 & 0.03 & 0.46 & 0.78 &  0.15 \\
PoL5& 43.03 & 50.06 & 0.79 & 0.36 & 0.63 & 0.40 & 0.001 \\
\hline
\end{tabular}
\caption{Same as table \ref{tbl1} but for the polynomial functions. $P_{GRB,rate}$ is the goodness of fit between our data 
and the observed GRB rate density.}
\label{tbl2}
\end{center}
\end{table}     

\section{REDSHIFT EVOLUTION ON THE NORMALIZATION AND SLOPE PARAMETERS} \label{redshift evolution}

\begin{figure}[htbp]
\centering
\includegraphics[width=0.49\hsize,angle=0,clip]{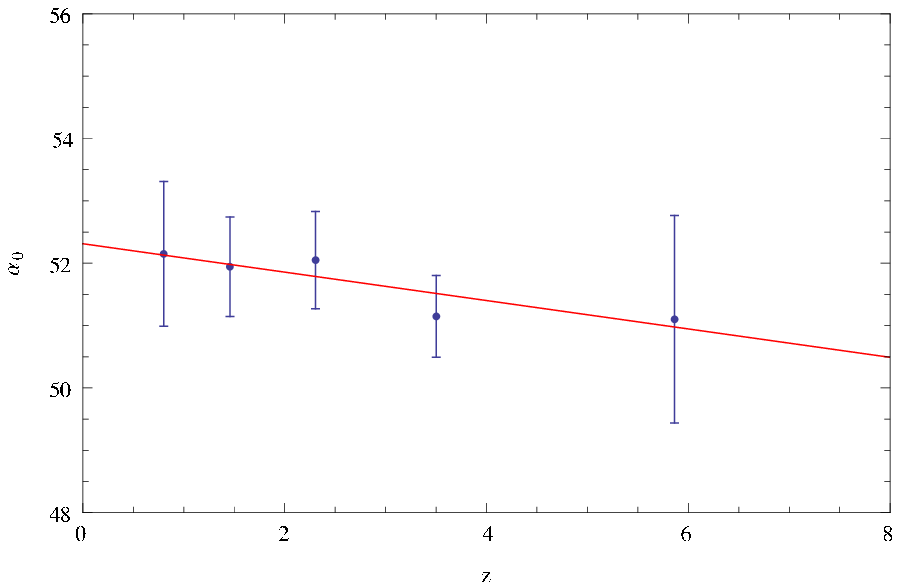}
\includegraphics[width=0.49\hsize,angle=0,clip]{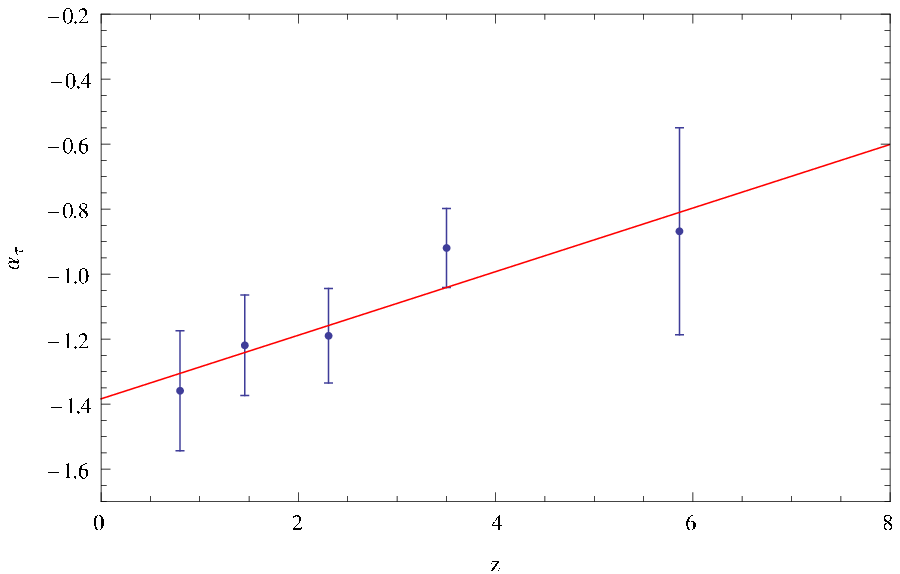}
\caption{\footnotesize $\alpha_{\tau}$ and normalization $\alpha_0$
using a linear function $\alpha_0=-0.22 x + 52.31$ (left panel) and $\alpha_{\tau}=0.10 x-1.38$ (right panel).}
   \label{fig:5}
\end{figure}

\begin{figure}[htbp]
\centering
\includegraphics[width=0.49\hsize,angle=0,clip]{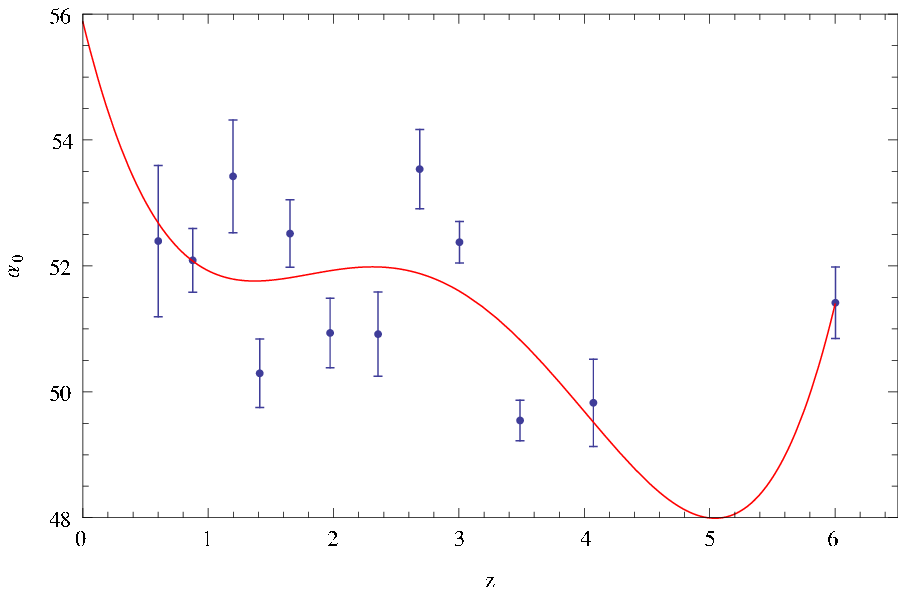}
\includegraphics[width=0.49\hsize,angle=0,clip]{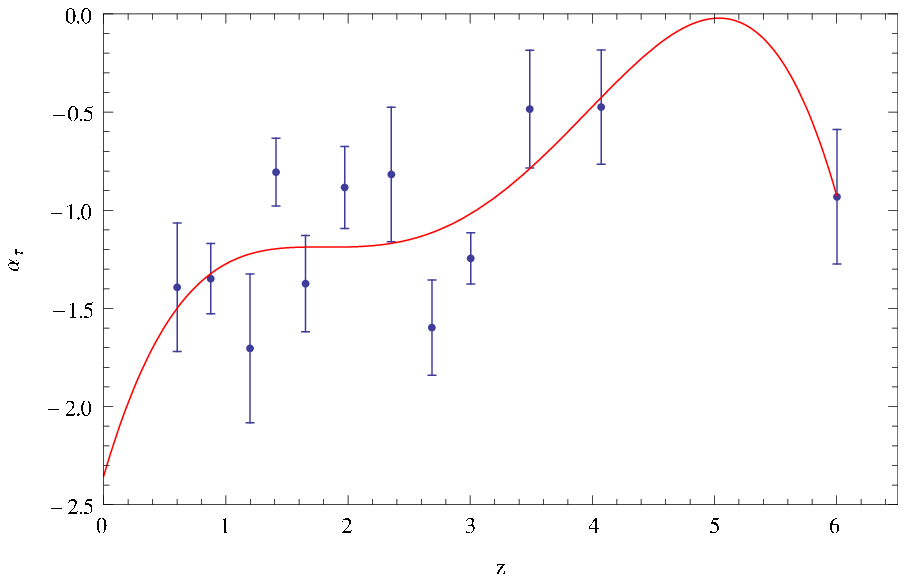}
\caption{\footnotesize $\alpha_{\tau}$ and normalization $\alpha_0$
using a polynomial function $\alpha_0=55.87 - 8.13 x + 5.53 x^2 - 1.48 x^3 + 0.13 x^4$ (left panel)
and $\alpha_{\tau}=-2.35 + 2.13 x - 1.39 x^2 + 0.37 x^3 - 0.03 x^4$ (right panel).}
   \label{fig:10}
\end{figure}

As we have already mentioned in the previous paragraph the polynomial and the linear model for the $\epsilon(\lambda)$ are unknown, then assumptions need to be made. We chose these forms, because both normalization and slope of the LT correlation depend on the redshift either with a polynomial or 
with a simple power law. Therefore, these choices for the selection functions take into account of this redshift dependence. 
Namely, we consider a model redshift dependence because of the corresponding dependence of both luminosity and time. This has been already shown in Dainotti et al. (2013a) and currently inthe middle panel of Fig. \ref{fig:4} for the updated data sample.
 To study the behavior of the redshift evolution we plot the slope and the 
normalization values versus the redshift. These are obtained from the average values for the data set 
divided into 5 bins, see figure \ref{fig:5} and into 12 bins, see figure \ref{fig:10}. As we can see from both figures 
\ref{fig:5} and \ref{fig:10} the normalization parameter $\alpha_0$ decreases as the redshift increases, while the 
slope parameter $\alpha_{\tau}$ shows the opposite trend.
 Goodness of the fits is given by the probability
$P=0.79$ for the data set divided in 5 bins and $P=0.87$ for the one divided into 12 bins for the linear case,
while for the polynomial model $P=0.99$ and $P=0.94$ for the data set divided in 5 and 12 bins respectively. These results show
that both polynomial and linear fit are possible.

\section{IMPACT OF SELECTION EFFECTS} \label{impact of selection effects}
 The simulated samples generated as described above are 
input to the same Bayesian fitting procedure we use with real
data. For each input ($\alpha_{\tau}$, $\alpha_{\zeta}$, $\alpha_0$, $\sigma_{int}$) parameters, we 
simulate $\sim 50$ GRBs sample setting $\mathcal{N}_{sim} = 200$, while the
number of observed GRBs depend on the efficiency function used. We fit these samples assuming no 
redshift evolution in Eq. \ref{scaling}, i.e. forcing $\alpha_{\zeta}=0$ in the fit so that, for each
simulated sample, the fitting procedure returns both the
best fit and the median and 68\% confidence range of the parameters 
($\alpha_{\tau}$, $\alpha_0$, $\sigma_{int}$). In order to investigate whether the
selection effects impact the recovery of the input scaling laws, we fit linear relations of the form:

\begin{equation}
x_f = ax_{inp} + b 
\label{xf}
\end{equation}

 where $x_{inp}$ is the input value and $x_f$ can be either the best
fit (denoted as $x_{bf}$) or the median $x_{fit}$ value. When fitting
the above linear relation, we use the $\chi^2$ minimization for
$x_{bf}$, while a weighted fit is performed for $x_{fit}$ with weights
$\omega_i = 1/\sigma_i^2$ where $\sigma_i$ is the symmetrized $1\sigma$ error. Note that the
label $i$ here runs over the simulations performed for each given efficiency function.

\begin{figure}[htbp]
\centering
\includegraphics[width=0.482\hsize,angle=0,clip]{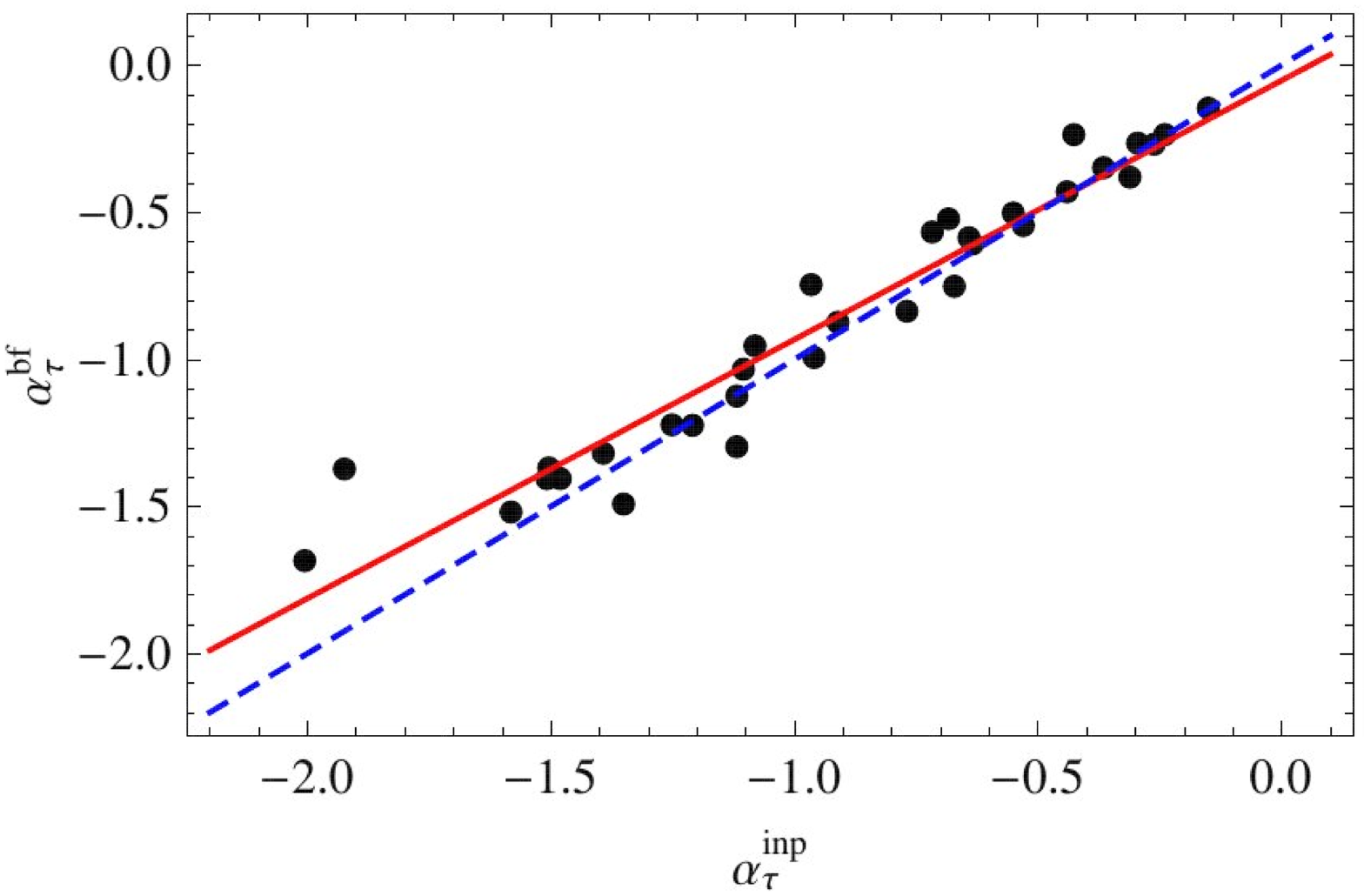}
\includegraphics[width=0.482\hsize,angle=0,clip]{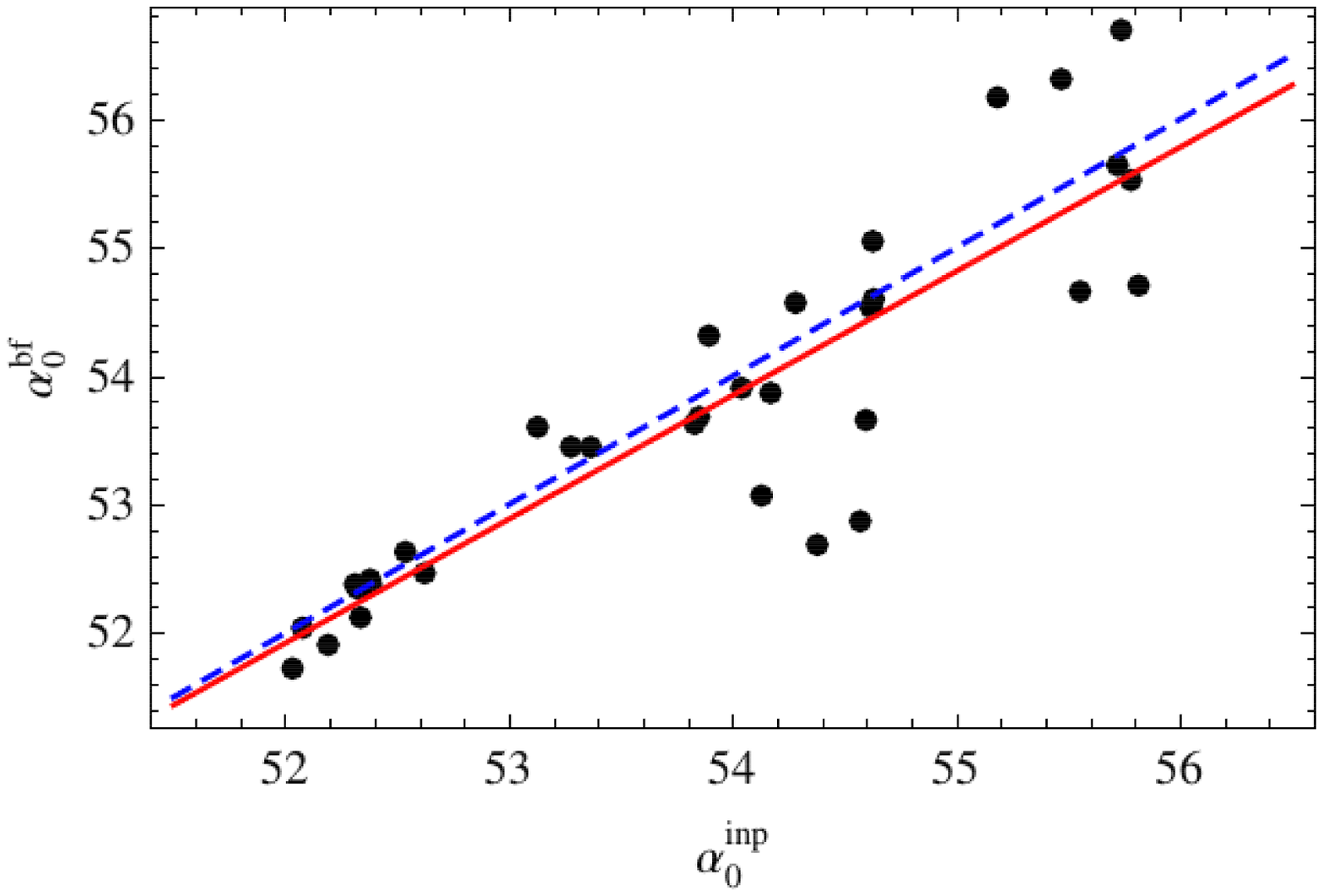}
\centering
\includegraphics[width=0.482\hsize,angle=0,clip]{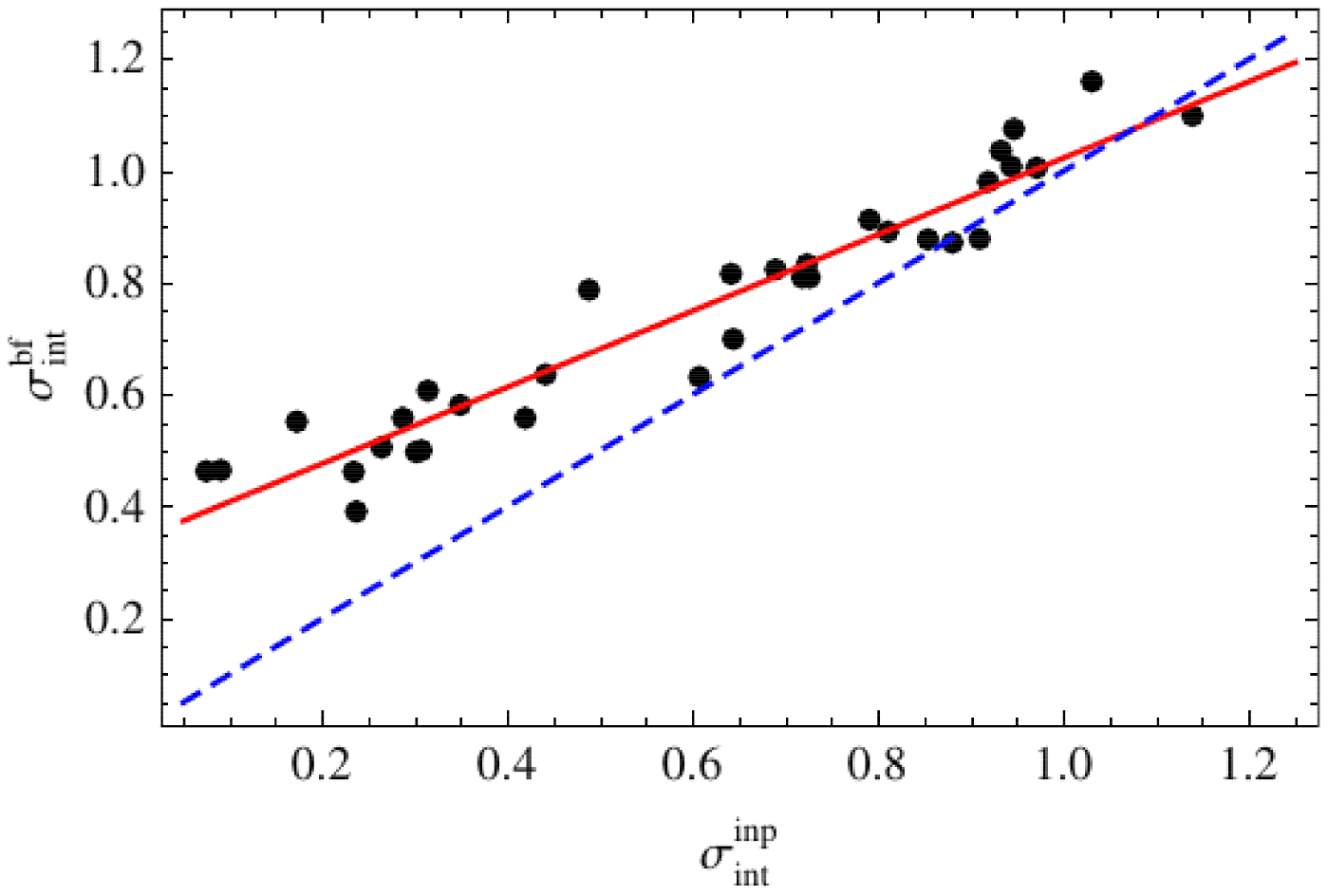}
\includegraphics[width=0.482\hsize,angle=0,clip]{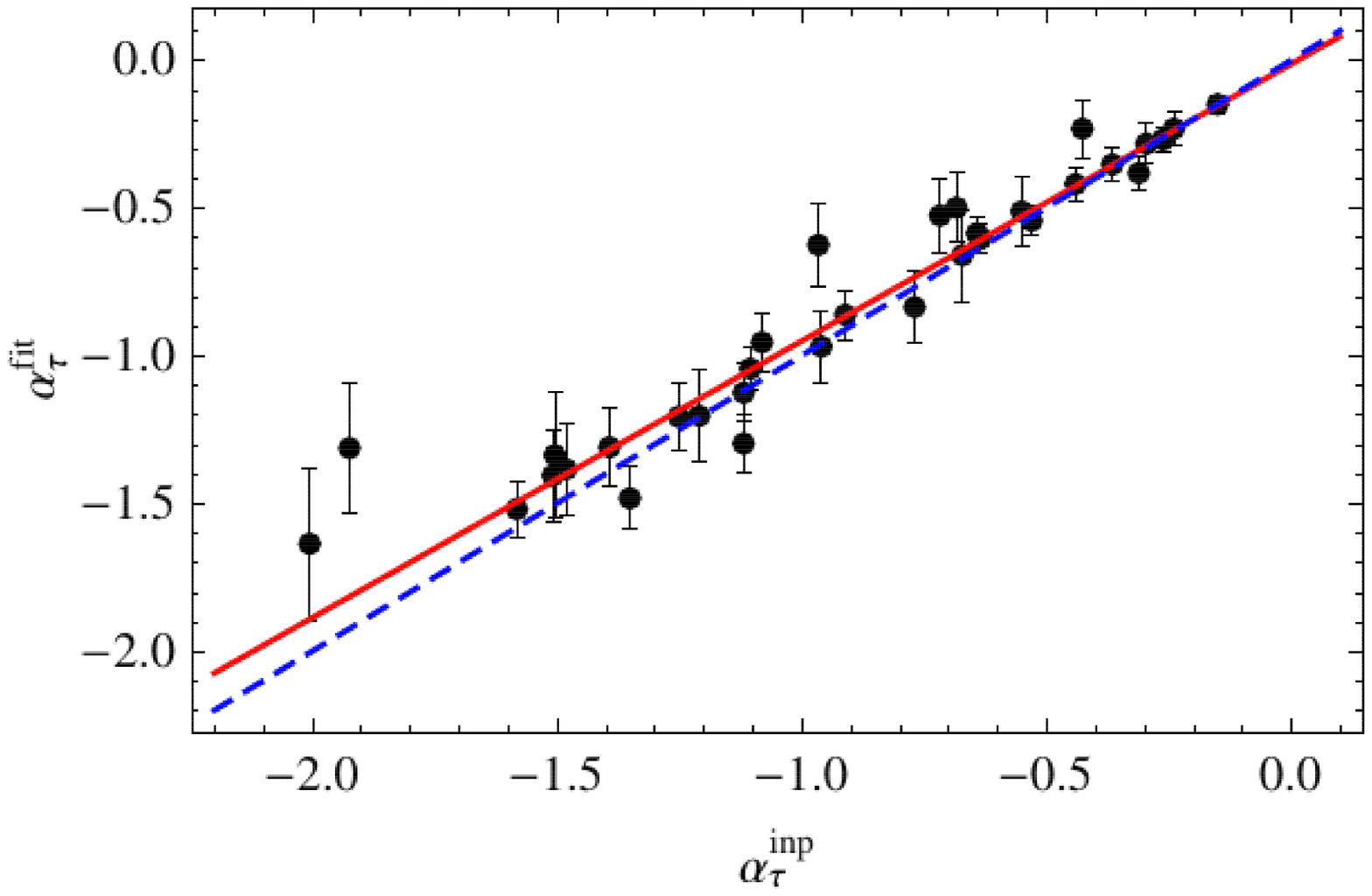}
\centering
\includegraphics[width=0.482\hsize,angle=0,clip]{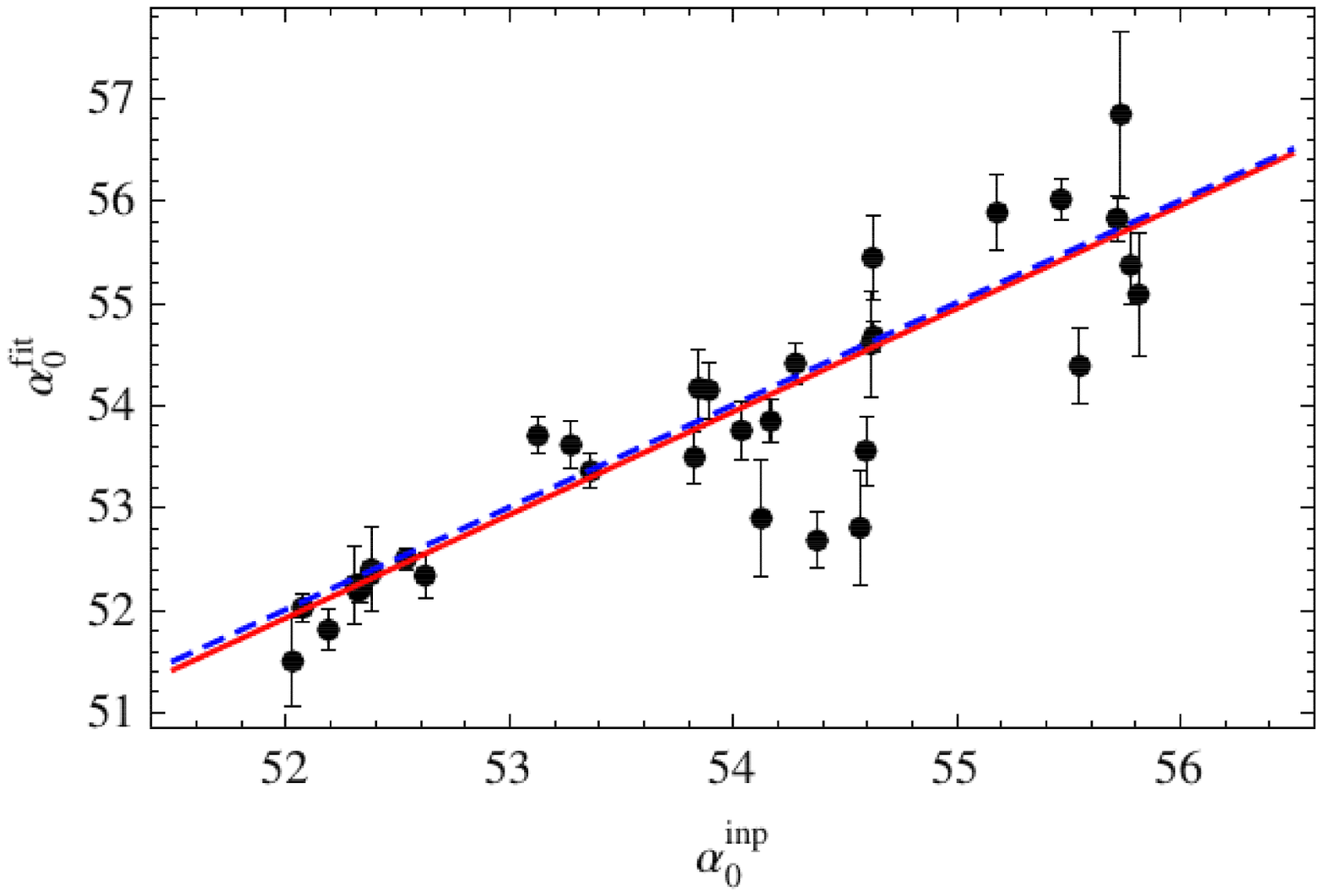}
\includegraphics[width=0.482\hsize,angle=0,clip]{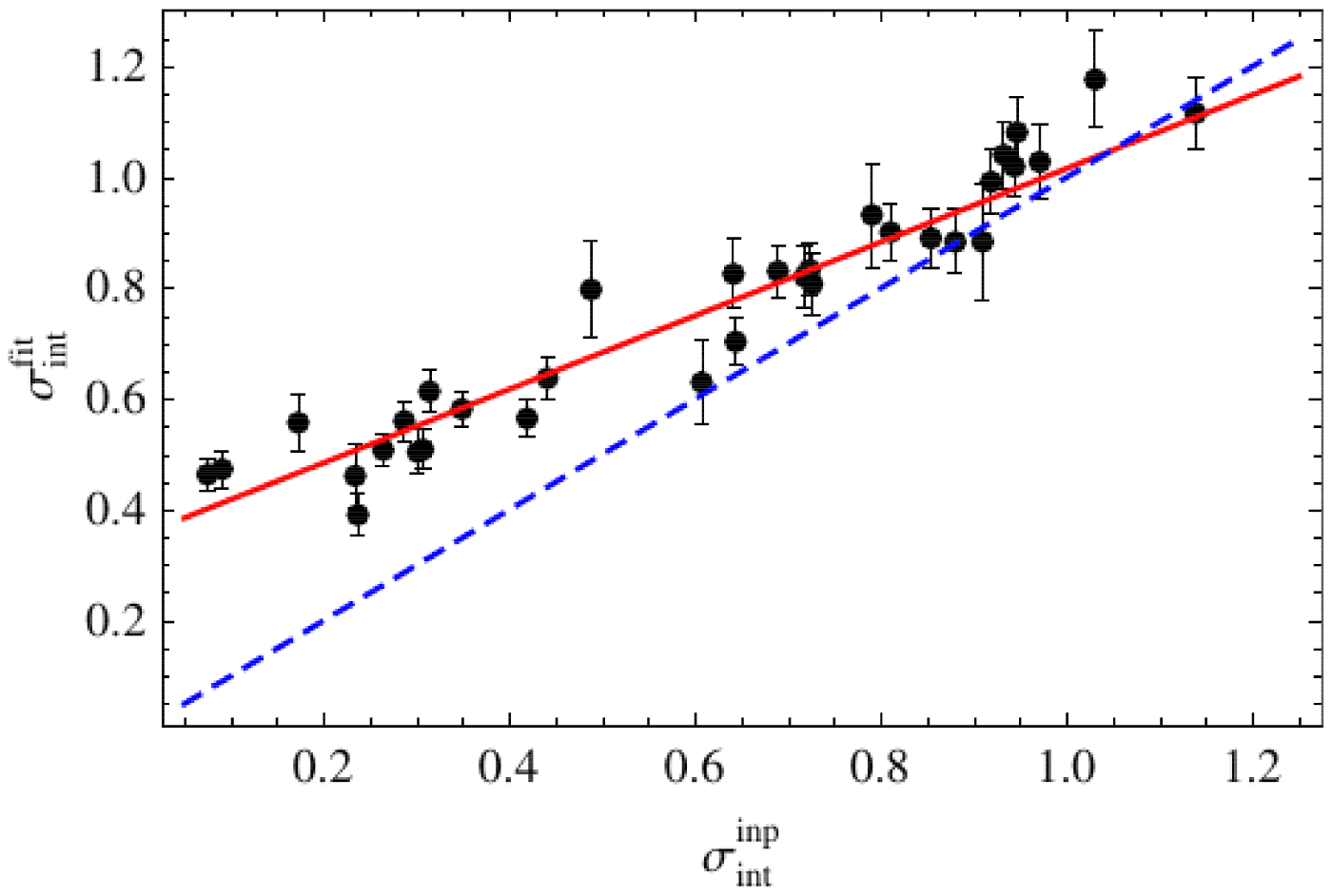}
\caption{\footnotesize Fitted vs input ($\alpha_{\tau}$, $\alpha_0$, $\sigma_{int}$) parameters obtained with the 
power law function. The first three panels refer to the best fit values, while the other three show the median 
values with the 1$\sigma$ error bars. Solid red line is the best fit line while blue dashed is the no bias line when 
$x_{inp}=x_f$.}
\label{fig:1}
\end{figure}

\begin{table}
\footnotesize
\begin{center}{
\begin{tabular}{cccccccc}
\hline
Id & $(a,b)^{\tau}_{bf}$ & $(a,b)^{\tau}_{fit}$ & $(a,b)^{\alpha_0}_{bf}$ & $(a,b)^{\alpha_0}_{fit}$ & $(a,b)^{\sigma}_{bf}$ & $(a,b)^{\sigma}_{fit}$ & $\frac{\Delta_x}{x}$ \\
\hline
PL1& (0.953,0.010) & (0.959,0.013) & (0.928,3.688) & (1.000,-0.073) & (0.593,0.354) & (0.616,0.355) & 0.004\\
PL2& (0.914,-0.008) & (0.873,-0.024) & (1.013,-0.836) & (0.989,0.292) & (0.689,0.299) & (0.643,0.341) & 0.002 \\
PL3& (0.880,-0.052) & (0.937,-0.013) & (0.965,1.729) & (1.008,-0.513) & (0.683,0.340) & (0.664,0.352) & 0.003\\
PL4& (0.946,0.024) & (0.964,0.024) & (0.995,-0.076) & (1.086,-4.905) & (0.614,0.364) & (0.585,0.380) & 0.006 \\
PL5& (0.916,-0.030) & (0.962,0.004) & (1.033,-2.067) & (0.828,9.095) & (0.716,0.333) & (0.679,0.356) & 0.005\\
\hline
\end{tabular}}
\caption{\footnotesize Slope $a$ and zero point $b$ of the fitted vs input parameters for both the best fit and median 
values (labeled with subscripts $bf$ and $fit$, respectively). The upperscript denotes the parameter fitted with 
($\tau$,$\alpha_0$,$\sigma$) referring to ($\alpha_{\tau}$, $\alpha_0$, $\sigma_{int}$), respectively. 
 $\frac{\Delta_x}{x}$ is the bias for each efficiency function considered.}{
\label{tbl3}} 
\end{center}
\end{table}     

\subsection{No redshift evolution}
 Here, we consider input models with $\alpha_{\zeta} = 0$, i.e. no redshift evolution of the scaling law (1).
It is worth noting that such an assumption is actually well motivated since it has been demonstrated in Dainotti et 
al. (2013a) that luminosity is almost not affected by redshift evolution, while time becomes to undergo redshift 
evolution for high redshift only. From our point of view, however, this case allows us to directly quantify the impact
of the efficiency functions on the recovery of the scaling correlation parameters since any deviation will only be due
to the selection effects and not to any attempt of compensating the missed evolution with z.

\subsubsection{No selection on $\tau$ ($\mathcal{E}_{\tau} = 1$)}
 We start by considering the idealized case of no selection of
$\tau$, i.e., we force $\mathcal{E}_{\tau} = 1$, and set the $\mathcal{E}_{\lambda}$ parameters as listed
in Tables \ref{tbl1} and \ref{tbl2} for the power-law and polynomial expressions, respectively. 
As an example, figure \ref{fig:1} shows the results for the efficiency function, while Table \ref{tbl3} summarizes
the (a, b) coefficients of the linear fit between the input and
recovered quantities. The closer is $a$ to 1, the less is the parameter biased, while $b \ne 0$ should not be taken as evidence for bias. This 
result is in perfect agreement with the intrinsic correlation slope, which is $-1.07^{+0.09}_{-0.14}$ \citep{Dainotti2013a}, when we consider as the best choice 
for the selection functions the ones that returns values of $a$ closer to 1.
If Equation \ref{xf} is fulfilled, we can estimate the relative bias as:

\begin{equation}
\frac{\Delta x}{x}=\frac{x_{inp}-x_f}{x_{inp}}=1-a-\frac{b}{x_{inp}},
\label{delta}
\end{equation}

\begin{table}
\footnotesize
\begin{center}{
\begin{tabular}{cccccccc}
\hline
Id & $(a,b)^{\tau}_{bf}$ & $(a,b)^{\tau}_{fit}$ & $(a,b)^{\alpha_0}_{bf}$ & $(a,b)^{\alpha_0}_{fit}$ & $(a,b)^{\sigma}_{bf}$ & $(a,b)^{\sigma}_{fit}$ & $\frac{\Delta_x}{x}$ \\
\hline
PoL1& (0.950,0.020) & (0.928,0.011) & (1.099,-5.545) & (1.178,-9.935) & (0.647,0.350) & (0.673,0.349) &  0.004\\
PoL2& (1.095,0.128) & (1.075,0.105) & (1.030,-1.662) & (1.023,-1.278) & (0.681,0.337) & (0.625,0.372) & 0.0008\\
PoL3& (0.984,0.063) & (0.936,0.015) & (0.711,15.258) & (0.988,0.376) & (0.741,0.289) & (0.685,0.336) &  0.007\\
PoL4& (0.870,-0.025) & (0.969,0.052) & (0.730,14.103) & (0.734,13.872) & (0.630,0.351) & (0.582,0.381) &  0.009\\
PoL5& (1.004,0.069) & (0.972,0.030) & (0.963,11.772) & (1.002,-0.374) & (0.581,0.384) & (0.549,0.402) &  0.18\\
\hline
\end{tabular}}
\caption{\footnotesize Same as Table \ref{tbl3} but for the polynomial $\mathcal{E}_{\lambda}$ model.}{
\label{tbl4}}
\end{center}
\end{table}     

 so that we can accept $b \ne 0$ if $x_{inp}$ is much larger than $b$. This is indeed the case for 
$x_{inp} = \alpha_0$ which takes typical values ($\sim 50$) much larger than the $b$ ones in Table \ref{tbl3}.\\ 
From the proximity between solid and dashed lines, which represent respectively the best fit line and the no bias line when $x_{inp}=x_f$, in the corresponding panels of Fig. \ref{fig:1}, we see that, for the power-law efficiency function (and no cut on $\tau$), 
both the slope and the zero point of the scaling relation are correctly recovered. 
The reason why is that the relative bias is negligible small notwithstanding the values 
of the parameters setting $\mathcal{E}_{\lambda}$. This is particularly true if one
relies on the median values as estimate since they are typically consistent 
with the no bias line within less than $2\sigma$.\\
 The above results have been obtained considering a
power-law $\mathcal{E}_{\lambda}$ so that it is worth investigating whether they
critically depend on this assumption. We have therefore repeated the analysis 
for the polynomial $\mathcal{E}_{\lambda}$ models in Table \ref{tbl2}
obtaining the results in Table \ref{tbl4}. A comparison with the values 
in Table \ref{tbl4} shows that the (a, b) coefficients are similar
so that one could preliminarily conclude that the shape of the
efficiency function does not play a major role in the 
determination of the bias. Actually, although the functional
expressions are different, both the power-law and the 
polynomial selection functions are qualitatively similar with $\mathcal{E}_{\lambda}$
increasing with $\lambda$ over a comparable range. Although such
a behaviour is likely common to any reasonable $\mathcal{E}_{\lambda}$, we can
not exclude a priori that non monotonic selection functions
do actually exist. What would the results be in such a case is
not clear so that we prefer to be cautious and conclude that the bias is roughly the same whichever 
monotonic $\mathcal{E}_{\lambda}(\lambda)$ function is used, but not for all the possible $\mathcal{E}_{\lambda}$ 
functions. For non-monotonic shape of selection function, see \cite{Stern2001}, in which
an assumed detection efficiency function, defined as the ratio of the number of detected test bursts to the number of 
test bursts applied to the data versus the expected peak count rate, is given by:

\begin{equation}
E(c_e)=0.70 \times [1-exp[-(\frac{c_e}{c_{e,0}})^2]]^{\nu},
\end{equation}

 where $c_{e,0}=0.097$ counts $s^{-1}$ $cm^{-2}$ and $\nu=2.34$ are two constants. However,
quoting from Stern et al. (2001), the best possible efficiency quality has still not yet been 
achieved because in fact the detection efficiency depends on the peak count rate rather then on the 
time-integrated signal.

\begin{table}
\footnotesize
\begin{center}
\begin{tabular}{cccccccc}
\hline
Id & $(a,b)^{\tau}_{bf}$ & $(a,b)^{\tau}_{fit}$ & $(a,b)^{\alpha_0}_{bf}$ & $(a,b)^{\alpha_0}_{fit}$ & $(a,b)^{\sigma}_{bf}$ & $(a,b)^{\sigma}_{fit}$ & $\frac{\Delta_x}{x}$\\
\hline
PLTa1& (0.842,-0.080) & (0.920,-0.016) & (1.099,-5.615) & (1.064,-3.650) & (0.615,0.363) & (0.585,0.387) & 0.04\\
PLTa2& (0.972,-0.016) & (0.963,-0.027) & (1.112,-5.980) & (1.148,-7.746) & (0.720,0.306) & (0.674,0.335) & 0.005\\
PLTa3& (0.960,-0.025) & (0.974,-0.005) & (0.985,0.791) & (1.022,-1.228) & (0.725,0.329) & (0.690,0.345) & 0.004\\
PLTa4& (0.877,-0.021) & (0.933,-0.005) & (1.005,-0.611) & (0.959,2.196) & (0.702,0.304) & (0.644,0.345) & 0.09\\
PLTa5& (0.860,-0.056) & (0.901,-0.026) & (0.911,4.592) & (0.838,8.655) & (0.687,0.338) & (0.613,0.382) & 0.06\\
PLTa6& (0.904,-0.010) & (0.919,-0.003) & (0.734,14.060) & (0.397,32.119) & (0.678,0.326) & (0.680,0.330) & 0.08\\
PLTa7& (0.915,-0.044) & (0.998,0.035) & (1.119,-6.577) & (1.119,-6.474) & (0.736,0.311) & (0.722,0.327) & 0.02\\
PLTa8& (0.967,-0.003) & (0.961,-0.017) & (1.027,-1.485) & (1.178,-9.515) & (0.731,0.287) & (0.705,0.313) & 0.03\\
PLTa9& (0.933,-0.005) & (0.904,-0.031) & (1.026,-1.530) & (0.804,10.289) & (0.641,0.357) & (0.670,0.352) & 0.06\\
PLTa10& (1.040,0.056) & (0.997,0.028) & (0.753,13.308) & (0.822,9.345) & (0.736,0.329) & (0.713,0.344) & 0.04\\
\hline
\end{tabular}
\caption{\footnotesize Same as Table \ref{tbl3} but for the selection cuts on both ($\tau$, $\lambda$) and power-law $\mathcal{E}_x$ functions.}{
\label{tbl5}}
\end{center}
\end{table}     

\subsubsection{Selection cuts on both $\tau$ and $\lambda$}
 We now consider the case where the total selection function
may be factorized as $\mathcal{E}(\tau, \lambda) = \mathcal{E}_{\tau}(\tau)\mathcal{E}_{\lambda}(\lambda)$ with both 
$\mathcal{E}_x(x)$ functions being given by power-law or fourth order polynomial expressions. 
We consider 10 different arbitrary choices for both cases.
Note that we have to increase $\mathcal{N}_{sim}$ to 300 in order to have $\mathcal{N}_{obs} = 80-100$ as for
the models discussed in the previous subsection.\\
 Table \ref{tbl5} gives the (a,b) coefficients for the different models considered. 
A comparison with Table \ref{tbl3} shows that, on average, the bias on the parameters is roughly the same with
the median values giving smaller deviations and significant
bias on $\sigma_{int}$ only. A more detailed analysis, however, shows
that, while, in the $\mathcal{E}_{\tau}=1$ case, biases larger than $5\%$ are
of the order of $10\%$. Namely, from Table \ref{tbl3} and \ref{tbl4} we show that the relative biases,
$\frac{\Delta_x}{x}$, both in the linear and the polynomial case, give very small values from $0.2\%$ to $0.9\%$, with
the only exception of $1$ polynomial function, in which the bias is $18\%$, thus giving $P=10\%$ of having larger bias
than $5\%$. If we consider selection cuts on both $\tau$ and $\lambda$ we notice in Table \ref{tbl5} that number of 
bias whose value is greater than $5\%$ are 4, thus increasing their probability to occur ($P\sim 40$\%).
This can be qualitatively explained noting that a cut on $\lambda$ only removes the points in the luminosity axis
thus possibly shifting the best fit relation, but not changing the slope. On the contrary, removing points
also along the horizontal $\tau$ axis can change the slope $\alpha_{\tau}$
too and hence also affects ($\alpha_0$, $\sigma_{int}$) because of the 
correlation among these parameters and $\alpha_{\tau}$. Similar results are
obtained when both $\mathcal{E}_x$ functions are modeled with fourth
order polynomials so that we will not discuss this case here.
We stress that when $\frac{\Delta_x}{x}$ are larger than $6\%$, than the slope of the correlation is farther
from $-1.0$ compared to cases in which $\frac{\Delta_x}{x} \le 0.06$. In fact, in the first case the slopes values 
range from $0.86$ to $0.91$, see the functions $PLTa4$ and $PLTa6$ in Table \ref{tbl5}. These values are not 
compatible in 1 $\sigma$ with the claimed intrinsic slope of the correlation, $-1.07^{+0.09}_{-0.14}$. If we consider,
instead, the lowest $\frac{\Delta_x}{x}$, then we obtain ranges of $a=(0.94,0.99)$ thus showing full compatibility in
1 $\sigma$ with the intrinsic slope. In this way we have quantitatively confirmed the existence of the $L_X-T^{*}_a$ 
correlation with the same intrinsic slope as in Dainotti et al. (2013a) if appropriate selection functions are chosen.

\section{Conclusions}\label{Conclusions}
 Here we built a general method to evaluate selection effects for GRB correlations not knowing a priori the 
efficiency function of the detector used. We have tested this method on the LT correlation. We chose a set of 
GRBs and assuming Gaussian distributions for the variables involved, for luminosity and time, and also a particular 
shape for the GRBs rate density. We simulated a mock sample of data in order to consider the selection 
effects of the detectors. As we can see in paragraph \ref{simulating the sample} assuming the correct observed
GRBs rate density shape was not an easy task. In fact, we explored different methods \citep{Li08,Robertson2012,kistler13,Hopkins2006} that use 
several SFR shapes to understand which one best matches the afterglow plateau data distribution including the selection functions, see Fig. \ref{fig:3}. The most reliable fits for the  GRB plateau data is the SFR
used by Li (2008), while the best efficiency functions for $\epsilon(\lambda)$ that match the GRB density rate can be both two polynomial and two linear, see Fig. \ref{fig:3}. Table \ref{tbl1} and \ref{tbl2} show the probability that the density rate fits the afterglow plateau GRB rate assuming those efficiency functions.
However, we assumed there could be selection effects both for luminosity and time. In 
particular, the bias is roughly the same whichever monotonic efficiency function for the luminosity 
detection $\mathcal{E}_{\lambda}$ is taken. From Table \ref{tbl3} and \ref{tbl4} we show that the relative 
biases, $\frac{\Delta_x}{x}$, both in the linear and the polynomial case, give very small values from $0.2\%$ to $0.9\%$, with
the only exception of $1$ polynomial function, in which the bias is $18\%$, thus giving $P=10\%$ of having larger bias
than $5\%$. If we consider selection cuts on both $\tau$ and $\lambda$ we notice in Table \ref{tbl5} that number of 
bias whose value is greater than $5\%$ are 4, thus increasing the probability of having such biases ($P\sim 40$\%).
 In addition, we studied selection effects in the LT correlation assuming also a combination of the 
luminosity and time detection efficiency functions. Different values for the parameters of the 
efficiency functions in the detectors are taken into account as described in the paragraph \ref{impact of selection effects}. This gives distinct fit values that inserted in Equation 
\ref{xf} allow to study the scattering of the correlation and its selection effects.
 We have quantitatively confirmed the existence of the $L_X-T^{*}_a$ correlation with the same intrinsic slope as in Dainotti et al. (2013a) if appropriate selection functions are chosen. In particular, when $\frac{\Delta_x}{x}$ are larger than $6\%$, than the slope of the correlation is farther
from $-1.0$ compared to cases in which $\frac{\Delta_x}{x} \le 0.06$. The lowest $\frac{\Delta_x}{x}$ leads to ranges of $a=(0.94,0.99)$ thus showing full compatibility in
1 $\sigma$ with the intrinsic slope.
Finally, the fact that the correlation is not generated by the biases themselves is a significant and further
step towards considering a set of GRBs as standard candles and their possible and useful application as a cosmological 
tool.

\section{Acknowledgements}
 This work made use of data supplied by the UK Swift Science Data Centre at the
University of Leicester. We are particularly grateful to Cardone, V.F. for the initial contribution to this work. M.G.D. and S.N. are grateful to the iTHES Group discussions at Riken.
M.D is grateful to the support from the JSPS Foundation, (No. 25.03786). N. S. is grateful to JSPS (No.24.02022, No.25.03018, No.25610056,  No.26287056) $\&$ MEXT(No.26105521), R. D.V. is grateful to 2012/04/A/ST9/00083.

\bibliography{biblio}{}

\end{document}